\documentclass[preprint,3p]{elsarticle}

\usepackage[utf8]{inputenc}
\usepackage{amsmath}
\usepackage{amsfonts}
\usepackage{amssymb}
\usepackage{graphicx}
\usepackage{hyperref}
\usepackage{braket}
\usepackage{multirow}
\usepackage{enumitem}
\usepackage{bbm}
\usepackage{units}
\usepackage[normalem]{ulem}
\usepackage{grffile}
\usepackage{xcolor}
\usepackage{tikz}
\usetikzlibrary{decorations.pathreplacing,calligraphy,backgrounds}

\newcommand{\olsi}[1]{\,\overline{\!{#1}}} 

\newcommand{\f}{f}
\newcommand{\cS}{\widetilde{\cal T}}  
\newcommand{\cK}{\olsi K}	
\newcommand{\cR}{{\cal T}}  
\newcommand{\cG}{{\cal K}}	
\newcommand{\cH}{\widetilde{\cal K}}	
\newcommand{\cT}{\overline T}	
\newcommand{\cV}{\widetilde V}	
\newcommand{\cc}{\tilde c}  
\newcommand{\cj}{\tilde j}  
\newcommand{\cf}{\tilde f}	
\newcommand{\cg}{\tilde g}	
\newcommand{\xx}{(x,x+\hat1)}
\newcommand{\X}{\cx}
\newcommand{\cx}{X}
\newcommand{\Mem}{\text{Mem}}
\newcommand{\Comp}{\text{Comp}}

\renewcommand{\j}{j}
\newcommand{\jp}{j'}
\newcommand{\clf}{\cj}

\newcommand{\SU}{\text{SU}}
\newcommand{\U}{\text{U}}
\def\sl3c{\text{SL}(3,\mathbb{C})}
\def\su3{\text{SU(3)}}

\newcommand{\J}{\boldsymbol{j}}
\newcommand{\dir}[1]{$\hat{#1}$}
\newcommand{\logZ}{\log Z}

\graphicspath{{plots/}}

\begin{document}

\title{Grassmann higher-order tensor renormalization group approach for two-dimensional strong-coupling QCD}

\author[1]{Jacques Bloch}
\ead{jacques.bloch@ur.de}
\author[1,2]{Robert Lohmayer}
\ead{robert.lohmayer@ur.de}

\address[1]{Institute for Theoretical Physics, University of Regensburg, 93040 Regensburg, Germany}
\address[2]{Leibniz Institute for Immunotherapy (LIT), 93053 Regensburg, Germany}

\date{\today}

\begin{abstract}
We present a tensor-network approach for two-dimensional strong-coupling QCD with staggered quarks at nonzero chemical potential. After integrating out the gauge fields at infinite coupling, the partition function can be written as a full contraction of a tensor network consisting of coupled local numeric and Grassmann tensors.  To evaluate the partition function and to compute observables, we develop a Grassmann higher-order tensor renormalization group method, specifically tailored for this model. 
During the coarsening procedure, the blocking of adjacent Grassmann tensors is performed analytically, and the total number of Grassmann variables in the tensor network is reduced by a factor of two at each coarsening step. The coarse-site numeric tensors are truncated using higher-order singular value decompositions.
The method is validated by comparing the partition function, the chiral condensate and the baryon density computed with the tensor method with exact analytical results on small lattices up to volumes of $4\times4$.
For larger volumes, we present first tensor results for the chiral condensate as a function of the mass and volume, and observe that the chiral symmetry is not broken dynamically in two dimensions. We also present tensor results for the number density as a function of the chemical potential, which hint at a first-order phase transition.
\end{abstract}

\maketitle

\section{Introduction}

The QCD phase diagram is a key research topic in modern particle physics, but its study with Monte Carlo methods in lattice QCD is hindered by the sign problem caused by the determinant of the Dirac operator, which becomes complex in the presence of a chemical potential $\mu$. Various methods developed to circumvent the sign problem, such as reweighting, Taylor expansion in $\mu$, analytic continuation from imaginary $\mu$, complex Langevin, thimbles and path optimization, have been applied to QCD, but none of these can successfully reach regimes where $\mu/T>1$. The method of dual variables shows some promise as it strongly reduces the sign problem, however, until now the dualization was mainly applied to the strong-coupling limit of QCD \cite{Rossi:1984cv,Karsch:1988zx,Fromm:2010lga,deForcrand:2009dh}. An attempt to go beyond this limit was made using the next-to-leading order term in the strong-coupling expansion \cite{Forcrand2014}. The worm algorithm \cite{Prokofiev:2001zz} is the method of choice to simulate QCD in the strong-coupling limit in its dual formulation.

As an alternative to Monte Carlo methods, tensor-network methods have recently been applied with success to various statistical systems. These methods can be categorized into Hamiltonian (or Hilbert-space) tensor methods and Lagrangian methods, which aim to compute the finite-temperature partition function. To study systems in thermal equilibrium we will constrain our discussion to the latter. Originally, the tensor renormalization group (TRG) method was proposed for two-dimensional systems \cite{Levin:2006jai}. This method was modified to be applicable to higher-dimensional systems in the higher-order tensor renormalization group (HOTRG) method \cite{Xie_2012}, which is based on the higher-order singular value decomposition (HOSVD) \cite{DeLathauwer2000}. The TRG and HOTRG methods have been applied to a variety of problems in classical and quantum statistical physics, such as spin systems or gauge systems in two, three and four dimensions. Even some systems with a complex action, i.e., with a sign problem, were successfully studied, as for example the three-dimensional O(2) model with a chemical potential \cite{Bloch:2021mjw}. For systems with fermions, which are represented by Grassmann variables in the partition function, the Grassmann HOTRG (GHOTRG) was recently developed for cases where the Grassmann variables cannot be integrated out locally \cite{Shimizu:2014uva,Takeda:2014vwa,Sakai:2017jwp}.

The aim of the current paper is to demonstrate the applicability of tensor-network methods to strong-coupling QCD with staggered quarks.
The HOTRG method cannot be applied as such to strong-coupling QCD as non-local sign factors occur in the meson-baryon-loop representation of the partition function \cite{Rossi:1984cv,Karsch:1988zx}. Clearly, this property is impossible to encode in a local tensor. 
To resolve this problem, we do not integrate out all Grassmann variables as in the meson-baryon-loop representation, but keep the baryonic combinations of Grassmann variables in a Grassmann tensor. Then, the partition function can be written as a full contraction of a tensor network with local numeric and Grassmann tensors. To evaluate the partition function we then apply an iterative blocking procedure, which uses ideas of the original GHOTRG method, but is specifically tailored for strong-coupling QCD.

We validate our Grassmann tensor-network method by comparing its results for the partition function, the chiral condensate and the number density with exact analytical results computed on small lattices of sizes up to $4\times 4$. Then, we apply the method to larger lattices to compute the chiral condensate as a function of mass and volume,  and observe that the chiral symmetry is not broken dynamically in the two-dimensional case. Furthermore, we compute the number density at nonzero chemical potential, which hints at a first-order phase transition. We also briefly discuss the convergence of the tensor-network results with increasing bond dimension.

The paper is structured as follows. In Sec.~\ref{sec:tensor} we reformulate the partition function of strong-coupling QCD as a tensor network of numeric and Grassmann tensors. In Sec.~\ref{sec:ghotrg} we introduce auxiliary Grassmann variables in order to decouple the nearest-neighbor interaction terms in the different directions. Then, we discuss how the lattice can be coarsened by blocking adjacent local tensors. We present our numerical results in Sec.~\ref{sec:results} and our  conclusions in Sec.~\ref{sec:conclusions}.

\section{Strong-coupling QCD and its tensor formulation}
\label{sec:tensor}

In the strong-coupling limit ($\beta\to 0$) of QCD, the gauge action vanishes and only the fermion action survives. For a single staggered quark field\footnote{In two dimensions, a single staggered quark field leads to two ``tastes'' in the continuum limit. Taste degrees of freedom are often interpreted as different physical flavors.} with mass $m$, the lattice action is
\begin{align}
S_F = \sum_x \left\{
\eta_{x,1} \gamma \bar\psi_x \left[e^{\mu} U_{x,1}\psi_{x+\hat1} - e^{-\mu} U^\dagger_{x-\hat1,1}\psi_{x-\hat1}\right] +  \eta_{x,2}\bar\psi_x \left[U_{x,2}\psi_{x+\hat2} - U^\dagger_{x-\hat2,2}\psi_{x-\hat2}\right] + 2m \bar\psi_x\psi_x 
\right\} ,
\end{align}
where $x\in\{1,\ldots,V\}$ enumerates the sites on a lattice with temporal extent $L_1$, spatial extent $L_2$, and volume $V=L_1 L_2$. For tensor-network studies $L_1$ and $L_2$ are taken to be powers of 2. The \SU(3) matrices $U_{x,\nu}$ are defined on the links of the lattice,
$\psi_x$ and $\bar\psi_x$ are 3-dimensional vectors of Grassmann variables, representing the colored quark and antiquark fields on the site $x$.
The staggered phases are $\eta_{x,1}=1$ and $\eta_{x,2} = (-1)^{x_1}$, where $x_1$ is the time coordinate of site $x$.
The quark chemical potential $\mu$ and an anisotropy factor $\gamma$ are introduced for the Euclidean time direction.\footnote{The anisotropy allows for a continuous variation of the temperature continuously \cite{Damgaard:1985np}.} To describe the system in thermal equilibrium, we use antiperiodic boundary conditions in the time direction and periodic boundary conditions in the space direction for the Grassmann variables.

In the infinite-coupling limit, the $\SU(3)$ gauge fields can be exactly integrated out \cite{Rossi:1984cv,Karsch:1988zx}, giving rise to a system of mesons and non-intersecting baryon loops. For each configuration contributing to the partition function, each lattice site is assigned either to a baryon loop or to a mesonic contribution, as all Grassmann variables must be saturated, i.e., each site has to contain 3 quarks and 3 anti-quarks, in order to contribute to the partition function. The partition function can then be written as \cite{Fromm:2010lga}
\begin{align}
Z = \int \left[\prod_x d\psi_x d\bar\psi_x\right] \prod_x e^{2mM_x} \prod_{\nu=1,2} z_{x,\nu} ,
\label{ZSQCD}
\end{align}
where the differentials are defined as
\begin{align}
d\psi_x d\bar\psi_x = d\psi_{x,3} d\psi_{x,2} d\psi_{x,1} d\bar\psi_{x,1} d\bar\psi_{x,2} d\bar\psi_{x,3} 
= d\psi_{x,1} d\bar\psi_{x,1} d\psi_{x,2} d\bar\psi_{x,2} d\psi_{x,3} d\bar\psi_{x,3}
\end{align}
and\footnote{Note that there is no back-and-forth baryonic contribution on the same link, as this is identical to the triple-meson contribution between two sites, which is already taken into account in the mesonic contribution.}
\begin{align}
z_{x,\nu} = 
\eta_{x,\nu} \zeta_\nu \bar B_x B_{x+\hat\nu} - \eta_{x,\nu} \zeta_{-\nu} \bar B_{x+\hat\nu} B_x
+
\sum_{k_{x,\nu}=0}^3 \frac{(3-k_{x,\nu})!}{3!k_{x,\nu}!}
\left((\eta_{x,\nu}\gamma^{\delta_{\nu,1}})^2 M_x M_{x+\hat\nu}\right)^{k_{x,\nu}}
\label{zxnu}
\end{align}
with mesonic combinations $M_x=\bar\psi_x\psi_x$, baryonic combinations $B_x =\frac{1}{3!}\epsilon_{i_1i_2i_3}\psi_{x,i_1}\psi_{x,i_2}\psi_{x,i_3}$, antibaryonic combinations $\bar B_x=\frac{1}{3!}\epsilon_{i_1i_2i_3}\bar\psi_{x,i_3}\bar\psi_{x,i_2}\bar\psi_{x,i_1}$ 
and
\begin{align}
\zeta_{\nu} 
= 
\begin{cases}
\gamma^3\exp(\pm 3\mu) & \text{for } \nu = \pm 1 \,, \\
1 & \text{else} \,.
\end{cases}
\end{align}

To integrate out the Grassmann variables, we first expand the exponential in the mass and write the product of sums in \eqref{ZSQCD} as a sum of products. Looking at a single term in the sum, i.e., a specific configuration, we observe that, for nonzero contributions, the sites have to be either baryonic or mesonic due to the Grassmann nature of the variables. Therefore the product over directions $\nu=1,2$ cannot mix baryonic and mesonic contributions on a single site.

To apply Monte Carlo simulations to strong-coupling QCD, the Grassmann variables in both the mesonic and baryonic combinations are integrated out.
The partition function then consists of configurations of closed, non-intersecting baryon loops with remaining sites saturated by meson contributions (including mass terms) \cite{Karsch:1988zx,Fromm:2010lga}.
A particularity of this representation is that each baryon loop contributes a multiplicative factor of $(-1)$ to the weight of the configuration to which it belongs, coming from a reordering of the Grassmann variables along the loop when performing the Grassmann integration.

Although the mesonic part of the partition function is easily converted into a consistent tensor-network formulation, as was already shown for $\U(N)$ \cite{Milde2021}, the baryonic contributions introduce a new problem as the baryon-loop sign factors are of a global nature and can therefore not be included in a local tensor without further ado. Therefore, HOTRG cannot be applied as such on this model. However, it turns out that this problem can be resolved using a variant of GHOTRG \cite{Shimizu:2014uva,Takeda:2014vwa,Sakai:2017jwp}, which we specifically develop for this model.

In the following we explicitly integrate out the Grassmann variables in the mesonic combinations, but leave the baryonic ones unintegrated to avoid the generation of non-local sign factors. After integration of the mesonic Grassmann combinations, the baryonic Grassmann variables $B_x$ and $\bar B_x$ can be regarded as fundamental (non-composite) Grassmann variables that are integrated over.
This results in\footnote{The integral sign denotes an integration over all Grassmann variables that occur for a given configuration $(\boldsymbol{k},\boldsymbol{l})$. In particular, when a configuration has no Grassmann contributions, our notation implies that there is no integral.}
\begin{align}
Z = \sum_{\boldsymbol{k},\boldsymbol{l}} \int \prod_x 
\Bigg\{
&\delta_{x\in{\cal B}}  \left[dB_x d\bar B_x \prod_{\nu=1,2} \eta_{x,\nu}^{|l_{x,\nu}|} \xi_\nu(l_{x,\nu}) (B_{x} \bar B_{x+\hat\nu})^{l_{x,\nu}^-} (\bar B_x B_{x+\hat\nu})^{l_{x,\nu}^+}  \right]
\notag\\
&+
\delta_{x\in{\cal M}} \left[h(n_x) \prod_{\nu=1,2} \alpha_\nu(k_{x,\nu})  \right]
\Bigg\} ,
\label{Z}
\end{align}
where the set of configurations  on a two-dimensional lattice of volume $V$ is the set of all tuples 
$\boldsymbol{k}=(k_{1,1},\dots,k_{V,2})$  and $\boldsymbol{l}=(l_{1,1},\dots,l_{V,2})$ 
of mesonic and net baryonic link occupation numbers $k_{x,\nu}\in\{0,1,2,3\}$ and $l_{x,\nu}\in\{-1,0,1\}$, respectively.
The occupation numbers $l^{\pm}_{x,\nu}$ for baryons and antibaryons are mutually exclusive in \eqref{Z}, see also \eqref{zxnu}, and can therefore be written as functions of the net occupation number $l_{x,\nu}$ with $l^{\pm}_{x,\nu} = l_{x,\nu}(l_{x,\nu}\pm1)/2\in\{0,1\}$, see also Table~\ref{table:l}.  
The weight functions in \eqref{Z} are
\begin{align}
\xi_\nu(l_{x,\nu}) &= \zeta_\nu^{l_{x,\nu}^+}\zeta_{-\nu}^{l_{x,\nu}^-} = 
\begin{cases}
\gamma^{3|l_{x,\nu}|}\exp(l_{x,\nu} 3\mu) & \text{if } \nu=1, \\
1 & \text{if } \nu=2,
\end{cases}
\\
\alpha_\nu(k_{x,\nu}) &= \frac{(3-k_{x,\nu})!}{3!k_{x,\nu}!}
\gamma^{2k_{x,\nu}\delta_{\nu,1}}, 
\\
h(n_x) &= \frac{3!}{n_x!}(2m)^{n_x} , \qquad
n_x = 3-\sum_\nu (k_{x,-\nu}+k_{x,\nu}), 
\label{hnx}
\end{align}
where we introduced the notation $k_{x,-\nu} \equiv k_{x-\hat\nu,\nu}$ and $l_{x,-\nu} \equiv l_{x-\hat\nu,\nu}$. 

\begin{table}
\centering
\begin{tabular}{|l||r|r|r|}
\hline
 $l$ & $-1$ & 0 & 1  \\
\hline
 $l^+$ & 0 & 0 & 1 \\
 $l^-$ & 1 & 0 & 0 \\
\hline
\end{tabular}
\caption{Relation between the occupation numbers $l$, $l^+$ and $l^-$ given by 
$l^{\pm} = l(l\pm1)/2$ and $l\in\{-1,0,1\}$.
}
\label{table:l}
\end{table}

For every configuration that yields a nonzero contribution to the partition function, each site $x$ is either baryonic or mesonic:
\begin{description}
\item[(a) baryonic site $x\in{\cal B}$:] All surrounding links must have $k=0$, i.e., $\sum_\nu (k_{x,-\nu}+k_{x,\nu})=0$. In order to yield a nonzero contribution to the partition function when all Grassmann variables are integrated out, each baryonic site $x$ must be occupied by exactly one factor of $B_x$ and one factor of $\bar B_x$. This means that for each baryonic site $x$ we require $l_{x,1}^{-}+l_{x,2}^{-}+l_{x,-1}^{+}+l_{x,-2}^{+}=1$ and $l_{x,1}^{+}+l_{x,2}^{+}+l_{x,-1}^{-}+l_{x,-2}^{-}=1$.
We represent the baryon condition by
\begin{align}
\delta_{x\in{\cal B}} =  \delta_{1,(l_{x,1}^{-}+l_{x,2}^{-}+l_{x,-1}^{+}+l_{x,-2}^{+})} \, \delta_{1,(l_{x,1}^{+}+l_{x,2}^{+}+l_{x,-1}^{-}+l_{x,-2}^{-})}
\prod_{\nu=1,2} \delta_{0,k_{x,-\nu}}\delta_{0,k_{x,\nu}} .
\label{xinB}
\end{align}

\item[(b) mesonic site $x\in{\cal M}$:] All surrounding links must have $l=0$ and the $\SU(3)$ condition requires $n_x\geq 0$, see \eqref{hnx}.
This is represented by the meson condition
\begin{align}
\delta_{x\in{\cal M}} =  \Theta(n_x) \prod_{\nu=1,2} \delta_{0,l_{x,-\nu}}\delta_{0,l_{x,\nu}} ,
\label{xinM}
\end{align}
where we use the convention $\Theta(0)=1$ for the Heaviside-theta function.

\end{description}

The configurations (index combinations) for which any single site $x$ is neither baryonic nor mesonic, i.e., with   indices such that $\delta_{x\in{\cal B}}=0$ and $\delta_{x\in{\cal M}}=0$, are not contributing to the partition function. 

Note that in the partition function \eqref{Z}, for each contributing index configuration $(\boldsymbol{k},\boldsymbol{l})$, the Grassmann variables appearing in the mesonic combinations have already been integrated out, and the remaining integrals only apply to the baryonic terms.
Because of the baryonic condition \eqref{xinB}, the Grassmann differentials for such a configuration can be rewritten as
\begin{align}
\prod_{x\in{\cal B}}  dB_x d\bar B_x
= \prod_x   (dB_x)^{l_{x,1}^{-}+l_{x,2}^{-}+l_{x,-1}^{+}+l_{x,-2}^{+}} (d\bar B_x)^{l_{x,1}^{+}+l_{x,2}^{+}+l_{x,-1}^{-}+l_{x,-2}^{-}} .
\label{DB}
\end{align}

For each configuration contributing to the partition function $\eqref{Z}$, any link is either mesonic ($k\neq0$, $l=0$), baryonic ($k=0$, $l\neq0$) or empty ($k=l=0$). Hence we can combine the mesonic and baryonic occupation numbers $k_{x,\nu}$ and $l_{x,\nu}$ into a single combined index $j_{x,\nu}$ of dimension 6, to reduce the total number of configurations in the partition function (i.e.,  we effectively reduce the number of configurations with zero weights).
The relation between the combined index $0\leq j \leq 5$ and the mesonic and baryonic occupation numbers $k$ and $l$ is given in Table~\ref{jkl}.

\begin{table}
\centering
\begin{tabular}{|l|c||c|c|c|c|c|c|}
\hline
combined index & $~j~$ & ~0~ & ~1~ & ~2~ & ~3~ & ~4~ & ~5~ \\
\hline
mesonic index & $k$ & 0 & 1 & 2 & 3 & 0 & 0\\
baryonic index & $l$ & 0 & 0 & 0 & 0 & $-1$ & $1$ \\
\hline
\end{tabular}
\caption{Mesonic and baryonic link occupation numbers $k$ and $l$ as a function of the combined index $j$.}
\label{jkl}
\end{table}

The partition function \eqref{Z} can be written as a full contraction of a tensor network where the local tensors have a numeric and a Grassmann part,
\begin{align}
Z = \sum_{\boldsymbol j}  \int \prod_x S^{(x)}_{j_{x,-1}j_{x,1}j_{x,-2}j_{x,2}} G^{(x)}_{l_{x,-1}l_{x,1}l_{x,-2}l_{x,2}} ,
\label{ZTN}
\end{align}
where each configuration is now characterized by its $2V$ indices $\J=(j_{1,1},\ldots,j_{V,2})$ and we again introduce the notation $j_{x,-\nu} \equiv j_{x-\hat\nu,\nu}$.
The local numeric tensors $S^{(x)}$ and the local Grassmann tensors $G^{(x)}$ have the following entries:
\begin{align}
S^{(x)}_{j_{x,-1}j_{x,1}j_{x,-2}j_{x,2}} &= 
\delta_{x\in{\cal B}}  \prod_{\nu=1,2}\eta_{x,\nu}^{|l_{x,\nu}|} \sqrt{\xi_{\nu}(l_{x,\nu})\xi_{\nu}(l_{x,-\nu})}
+ 
\delta_{x\in{\cal M}} \, h(n_x) \, \prod_{\nu=1,2} \sqrt{\alpha_{\nu}(k_{x,\nu})\alpha_{\nu}(k_{x,-\nu})} 
\,,\label{Tm}\\
G^{(x)}_{l_{x,-1}l_{x,1}l_{x,-2}l_{x,2}} &=  (dB_x)^{l_{x,1}^{-}+l_{x,2}^{-}+l_{x,-1}^{+}+l_{x,-2}^{+}} (d\bar B_x)^{l_{x,1}^{+}+l_{x,2}^{+}+l_{x,-1}^{-}+l_{x,-2}^{-}} \prod_{\nu=1,2}  (B_{x} \bar B_{x+\hat\nu})^{l_{x,\nu}^-} (\bar B_x B_{x+\hat\nu})^{l_{x,\nu}^+} ,
\label{Gm}
\end{align}
where the indices $k_{x,\nu}$ and $l_{x,\nu}$ are implicitly defined as functions of $j_{x,\nu}$ as in Table~\ref{jkl}, and we recall that $l_{x,\nu}^{\pm} = l_{x,\nu}(l_{x,\nu}\pm1)/2 \in \{0,1\}$, see Table~\ref{table:l}.

Note that each configuration $\J$ in the partition function \eqref{ZTN} selects one entry for each local tensor, and due to \eqref{Tm}, the nonzero tensor entries of $S$ can be classified as either ``mesonic" or ``baryonic".
For the mesonic entries, the corresponding entries of the Grassmann tensor are all equal to 1, as all $l^+$ and $l^-$ are zero around a mesonic site. For the baryonic entries, the corresponding entries of the Grassmann tensor are non-trivial and will play a crucial role in GHOTRG.\footnote{In \eqref{ZTN}-\eqref{Gm} we use the convention that $S^{(x)}G^{(x)}$ is zero whenever $S^{(x)}$ is zero. This is relevant when the indices of $G^{(x)}$ are such that $dB$ or $d\bar B$ would have powers different from 0 or 1.}

As all the interaction terms in $G^{(x)}$ involve an even number of Grassmann variables, they are mutually commuting and so their order does not matter. Furthermore, due to the factors $\delta_{x\in{\cal B}}$ and $\delta_{x\in{\cal M}}$ in $S^{(x)}$, the Grassmann tensors $G^{(x)}$ can be considered to be commuting with each other in \eqref{ZTN}, since for every nonzero entry of $S^{(x)}$ the corresponding entry of $G^{(x)}$ is Grassmann-even.

The local numeric tensor $S^{(x)}$ only depends on the site $x$ through the staggered phase $\eta_{x,\nu}$. As $\eta_{x,1}=1$ and $\eta_{x,2}=(-1)^{x_1}$, there are only two different realizations of $S^{(x)}$, for sites with odd and even time coordinates $x_1$, respectively.\footnote{To avoid multiple definitions of the local tensor that just differ in the staggered phase, the latter is not included explicitly in the computer implementation of the initial local tensor \eqref{Tm}, but is instead taken care of explicitly in the very first contraction, which is performed in the \dir1-direction, see \ref{app:staggered}.} The bond dimension of the initial local tensor is $D_\text{initial}=6$, corresponding to the dimension of the index $j$ in Table~\ref{jkl}. During the iterative blocking procedure, the bond dimensions of the coarse-lattice tensors, which would in principle grow exponentially, are truncated to a chosen value $D$ using HOSVD approximations \cite{DeLathauwer2000}.

In order to validate our version of the GHOTRG method, we will also investigate a simplified partition function containing only baryons. The sum over $\J$ in \eqref{ZTN} is then restricted such that all sites are baryonic, i.e.,  $x\in{\cal B}$ for all $x$.

\section{\boldmath Grassmann HOTRG for strong-coupling QCD}
\label{sec:ghotrg}

We now explain how to evaluate the partition function given by the full contraction of the Grassmann tensor network in \eqref{ZTN}.
The method can be summarized as being an iterative blocking procedure where each blocking step consists of two parts: First, new Grassmann tensors are generated on the coarse lattice, which reduces the number of Grassmann variables by a factor of two and gives rise to local sign factors. Then, an HOSVD approximation is applied to the contraction of two adjacent numeric tensors. In this process, the local sign factors are absorbed in the new numeric tensors on the coarse lattice.

The peculiarities of the strong-coupling QCD model, i.e., the use of staggered quarks and the existence of both, mesonic and baryonic contributions, require the development of a tailor-made GHOTRG.

\subsection{Decoupling the Grassmann interaction terms through auxiliary variables}

In order to integrate out the Grassmann variables $B_{x}$ and $\bar B_{x}$ in the partition function \eqref{ZTN} for one particular configuration $\J$, satisfying $x\in{\cal B}$ or $x\in{\cal M}$ for all $x$,
we decouple the interaction terms in different directions. This is achieved by introducing auxiliary Grassmann variables $c$ and inserting identities of the form
\begin{align}
\int \left(dc\right)^l \left(c\right)^l \equiv \left(\int dc\,c\right)^l=1 \quad \text{for $l=0,1$} .
\label{intaux}
\end{align}
The interaction terms $(\bar B_{x}B_{x+\hat\nu})^{l_{x,\nu}^+}$ and $(B_{x}\bar B_{x+\hat\nu})^{l_{x,\nu}^-}$ are mutually exclusive, i.e., $l_{x,\nu}^+$ and $l_{x,\nu}^-$ cannot simultaneously be equal to one, see Table~\ref{table:l}. Therefore, we can use the same auxiliary variable $c_{x,\nu}$ to rewrite the two interaction terms as
\begin{align}
\begin{aligned}
(\bar B_{x}B_{x+\hat\nu})^{l_{x,\nu}^+} &= 
\left(\bar B_{x}B_{x+\hat\nu} \int dc_{x,\nu} c_{x,\nu}\right)^{l_{x,\nu}^+}  
= \int (\bar B_{x}c_{x,\nu})^{l_{x,\nu}^+} (B_{x+\hat\nu}dc_{x,\nu})^{l_{x,\nu}^+}, \\
(B_{x}\bar B_{x+\hat\nu})^{l_{x,\nu}^-} &= 
\left(B_{x}\bar B_{x+\hat\nu} \int d c_{x,\nu}  c_{x,\nu}\right)^{l_{x,\nu}^-} 
= \int (B_{x} c_{x,\nu} )^{l_{x,\nu}^-}(\bar B_{x+\hat\nu}dc_{x,\nu})^{l_{x,\nu}^-},
\end{aligned}
\label{auxfields}
\end{align}
where each interaction term is split into two commuting factors. These identities can be applied for all $x$ and $\nu=1,2$ independently. 
The order of the Grassmann variables on the right hand side is chosen to facilitate the integration of $B_x$ and $\bar B_x$ below.
For later convenience we will also introduce the notation $c_{x,-\nu} \equiv c_{x-\hat\nu,\nu}$.

After introducing the auxiliary Grassmann variables using \eqref{auxfields}, all original Grassmann variables $B_x$ and $\bar B_x$ can be integrated out independently for different sites. To this end, we gather all eight interaction terms involving $B_x$ or $\bar B_x$ for one particular site $x$, together with the differentials contained in $G^{(x)}$ (the commuting pairs are reordered to gather the contributions in $B$ and $\bar B$ separately, such that the Grassmann integrations can be performed without generating additional sign factors). For $x\in{\cal B}$ or $x\in{\cal M}$, satisfying the conditions \eqref{xinB} and \eqref{xinM}, respectively, we obtain
\begin{align}
H^{(x)}_{l_{x,-1}l_{x,1}l_{x,-2}l_{x,2}} &= \int_{B_x} (dB_x)^{l_{x,1}^{-}+l_{x,2}^{-}+l_{x,-1}^{+}+l_{x,-2}^{+}}
(B_{x} c_{x,1})^{l_{x,1}^-}
(B_{x} c_{x,2})^{l_{x,2}^-}
(B_{x}dc_{x,-1})^{l_{x,-1}^+}
(B_{x}dc_{x,-2})^{l_{x,-2}^+} 
\notag\\&\times
\int_{\bar B_x} (d\bar B_x)^{l_{x,1}^{+}+l_{x,2}^{+}+l_{x,-1}^{-}+l_{x,-2}^-}
(\bar B_{x}c_{x,1})^{l_{x,1}^+}
(\bar B_{x}c_{x,2})^{l_{x,2}^+}
(\bar B_{x}d c_{x,-1})^{l_{x,-1}^-}
(\bar B_{x}d c_{x,-2})^{l_{x,-2}^-}
\notag\\
&= \int_{B_x} ( dB_x B_{x})^{l_{x,1}^{-}+l_{x,2}^{-}+l_{x,-1}^{+}+l_{x,-2}^{+}}
(c_{x,1})^{l_{x,1}^-} (c_{x,2})^{l_{x,2}^-}(dc_{x,-1})^{l_{x,-1}^+}(dc_{x,-2})^{l_{x,-2}^+}
\notag\\
&\times \int_{\bar B_x} (d\bar B_x \bar B_{x})^{l_{x,1}^{+}+l_{x,2}^{+}+l_{x,-1}^{-}+l_{x,-2}^-}
(c_{x,1})^{l_{x,1}^+}(c_{x,2})^{l_{x,2}^+}(d c_{x,-1})^{l_{x,-1}^-}(d c_{x,-2})^{l_{x,-2}^-}
\notag\\
&= \left[(c_{x,1})^{l_{x,1}^-} (c_{x,2})^{l_{x,2}^-}(dc_{x,-1})^{l_{x,-1}^+}(dc_{x,-2})^{l_{x,-2}^+}\right]
\left[(c_{x,1})^{l_{x,1}^+}(c_{x,2})^{l_{x,2}^+}(d c_{x,-1})^{l_{x,-1}^-}(d c_{x,-2})^{l_{x,-2}^-}\right]
\label{intB}
\end{align}
with $\int_B$ and $\int_{\bar B}$ indicating that we only integrate over the Grassmann variables $B$ and $\bar B$. 
If the chosen site $x$ is baryonic, then $H^{(x)}$ always contains exactly two Grassmann variables due to \eqref{xinB}, one from each product in square brackets. On the other hand, if $x$ is mesonic, $H^{(x)}=1$ since all $l^{\pm}_{x,\pm\nu}=0$. In both cases $H^{(x)}$ is Grassmann even.

Note that to collect all interaction terms involving $B_x$ and $\bar B_x$ in $Z$, resulting in \eqref{intB}, one also needs \eqref{auxfields} with $x\to x-\hat\nu$. When $x$ is on the lower edge of the lattice in the \dir\nu-direction, the interaction terms between $x-\hat\nu$ and $x$ will wrap around the lower edge of the lattice. These terms actually stem from the interaction terms in \eqref{ZTN} which wrap around the lattice at its upper edge in that direction (because \eqref{ZTN} only contains interactions from $x$ to $x+\hat\nu$ for $x\in\{1,\ldots,V\}$). For these interaction terms, the variables $B_x$ and $\bar B_x$ will be subjected to the boundary conditions in the direction \dir\nu. However, we always have
\begin{align}
\bar B_{y} B_{y'} = \bar B_{y \pm L_\nu\hat\nu} B_{y' \pm L_\nu\hat\nu} , \quad \text{for all $y,y',\nu$} 
\end{align}
since we use antiperiodic (for $\nu=1$) and periodic (for $\nu=2$) boundary conditions.
In order to preserve this property for products of original and auxiliary variables, appearing on the right hand side of \eqref{auxfields},
and to avoid explicit sign factors in the partition function, we choose the boundary conditions of the new auxiliary variables such that we always have
\begin{align}
a_{y}b_{y'} &= a_{y\pm L_\nu\hat\nu}b_{y'\pm L_\nu\hat\nu}
, \quad \text{for all $y,y',\nu$} ,
\end{align}
where $a_y,b_y$ are place holders for any of the original or auxiliary variables (or differentials).
The conditions above are automatically satisfied by requiring the auxiliary variables to satisfy the same boundary conditions as the original Grassmann variables $B$ and $\bar B$.

After integrating out $B_x$ and $\bar B_x$, according to \eqref{intB}, for all sites $x$ in the partition function \eqref{ZTN},
we are left with $2V$ new Grassmann variables $c_{x,\nu}$ and their differentials, and the partition function can be written as
\begin{align}
Z = \sum_{\J}  \int \prod_x S^{(x)}_{j_{x,-1}j_{x,1}j_{x,-2}j_{x,2}} H^{(x)}_{l_{x,-1}l_{x,1}l_{x,-2}l_{x,2}}
\label{ZH}
\end{align}
with numerical tensors $S^{(x)}$ and Grassmann tensors $H^{(x)}$ given in \eqref{Tm} and \eqref{intB}, respectively.
Recall that the indices $l_{x,\nu}$ of $H^{(x)}$ are implicit functions of $j_{x,\nu}$ as given in Table~\ref{jkl}.
For each configuration $\boldsymbol j$, the integral in \eqref{ZH} applies to all auxiliary fields having differentials with unit exponent.

The main difference compared to the original formulation \eqref{ZTN} for $Z$ is that the new Grassmann variables live on the links, while the original Grassmann variables where defined on the sites of the lattice. This is crucial for deriving a consistent blocking procedure, as will be shown in the next sections.

As explained above, the Grassmann tensors $H^{(x)}$ in the partition function can be considered to be commuting since every entry $H^{(x)}_{l_{x,-1}l_{x,1}l_{x,-2}l_{x,2}}$ is accompanied by a factor $\delta_{x\in\cal B}$ or $\delta_{x\in\cal M}$ in the numeric tensor $S^{(x)}$. 

To facilitate the further manipulations, we reorder the factors in $H^{(x)}$ in a canonical order (chosen such that the Grassmann integrations can be more easily performed when blocking two tensors as described in further sections),
\begin{align}
H^{(x)}_{l_{x,-1}l_{x,1}l_{x,-2}l_{x,2}} 
&= \omega_{l_{x,-1}l_{x,1}l_{x,-2}l_{x,2}}
(c_{x,1})^{l_{x,1}^{-}}
(c_{x,1})^{l_{x,1}^{+}} 
(c_{x,2})^{l_{x,2}^{-}}  
(c_{x,2})^{l_{x,2}^{+}} 
\notag
\\
&\hspace{2ex}\times 
(dc_{x,-2})^{l_{x,-2}^{+}}
(dc_{x,-2})^{l_{x,-2}^{-}}
(dc_{x,-1})^{l_{x,-1}^{+}}
(dc_{x,-1})^{l_{x,-1}^{-}}
\label{Kxlpm}
\end{align}
with sign factor\footnote{Note that for Grassmann variables $\psi$ and $\chi$, we can write $\psi^a \chi^b = (-1)^{ab} \chi^b\psi^a$ for $a,b=0,1$. The sign factor $\omega$ is simplified using the baryon condition \eqref{xinB} and the fact that $l^+ l^- = 0$ and $(l^\pm)^2=l^\pm$. The expression for $\omega$ is also valid for $x\in{\cal M}$, where $H^{(x)}=1$.}
\begin{align}
\omega_{l_{x,-1}l_{x,1}l_{x,-2}l_{x,2}} = (-1)^{l_{x,1}^{+} + l_{x,2}^{+} + l_{x,2}^{+}l_{x,1}^- + l_{x,-2}^{-}l_{x,-1}^{+}} ,
\label{sx}
\end{align}
which will  eventually be absorbed in the numeric tensor.
As $l^{\pm}_{x,\nu} = l_{x,\nu}(l_{x,\nu}\pm1)/2$, the products in \eqref{Kxlpm} can be rewritten as
\begin{align}
(c_{x,\nu})^{l_{x,\nu}^{+}} (c_{x,\nu})^{l_{x,\nu}^{-}}
= (c_{x,\nu})^{l_{x,\nu}^2} ,
\end{align}
such that the Grassmann tensor $H^{(x)}$ becomes
\begin{align}
H^{(x)}_{l_{x,-1}l_{x,1}l_{x,-2}l_{x,2}} 
&= \omega_{l_{x,-1}l_{x,1}l_{x,-2}l_{x,2}}
(c_{x,1})^{l_{x,1}^2} 
(c_{x,2})^{l_{x,2}^2} 
(dc_{x,-2})^{l_{x,-2}^2}
(dc_{x,-1})^{l_{x,-1}^2}  .
\label{HK}
\end{align}
We now introduce the notation $f_{x,\nu}\equiv f_{x,\nu}(j_{x,\nu})=l_{x,\nu}^2 \in \{0,1\}$ and identify $f_{x,-\nu}=f_{x-\hat\nu,\nu}$. After defining a new Grassmann tensor
\begin{align}
K^{(x)}_{f_{x,-1}f_{x,1}f_{x,-2}f_{x,2}}  = 
(c_{x,1})^{\f_{x,1}} 
(c_{x,2})^{\f_{x,2}} 
(dc_{x,-2})^{\f_{x,-2}}
(dc_{x,-1})^{\f_{x,-1}} 
\label{Kx}
\end{align}
and absorbing the sign factor $\omega$ in a new numeric tensor
\begin{align}
T^{(x)}_{\j_{x,-1}\j_{x,1}\j_{x,-2}\j_{x,2}} 
&=
\omega_{l_{x,-1}l_{x,1}l_{x,-2}l_{x,2}} S^{(x)}_{j_{x,-1}j_{x,1}j_{x,-2}j_{x,2}} ,
\label{initT}
\end{align}
the partition function \eqref{ZH} becomes
\begin{align}
Z = \sum_{\boldsymbol j} \int \prod_x
T^{(x)}_{\j_{x,-1}\j_{x,1}\j_{x,-2}\j_{x,2}} 
K^{(x)}_{f_{x,-1}f_{x,1}f_{x,-2}f_{x,2}} .
\label{ZTNjf}
\end{align}

Note that the Grassmann tensors $K^{(x)}$ can be considered to be commuting (Grassmann-even) in \eqref{ZTNjf}, 
since the baryonic condition $\delta_{x\in{\cal B}}$ included in $T^{(x)}$ implies $\f_{x,1}+\f_{x,2}+\f_{x,-1}+\f_{x,-2}=2$, see \eqref{xinB}, while the mesonic condition $\delta_{x\in{\cal M}}$ in $T^{(x)}$ implies $\f_{x,1}+\f_{x,2}+\f_{x,-1}+\f_{x,-2}=0$, see \eqref{xinM}.
Hence, the Grassmann tensors can be reordered in the partition function when performing the coarsening steps discussed below, without generating additional sign factors.

In the next sections we will describe the renormalization group (RG) steps, which coarsen the lattice iteratively and halve the number of lattice sites at each iteration. 
The blocking of two adjacent Grassmann tensors $K^{(x)}$ and $K^{(x+\hat\nu)}$ will produce a new tensor on the coarse lattice with a Grassmann structure identical to that of the original tensors, and a sign factor that can be absorbed in the coarse-lattice numeric tensor.
In the following we will call the index $f_{x,\nu}\equiv f_{x,\nu}(\j_{x,\nu})\in\{0,1\}$ the Grassmann parity of the index $j_{x,\nu}$. The Grassmann parity $f_{x,\nu}$ is the exponent of the Grassmann variable living on the link between $x$ and $x+\hat\nu$, see \eqref{Kx}.
In the original local Grassmann tensor $K^{(x)}_{f_{x,-1}f_{x,1}f_{x,-2}f_{x,2}}$, the indices $f=f(j)$ are related to $j$ by Table~\ref{jkl} and $f=l^2$. However, in general, the index $f$ is a function of the index $j$, which will be updated at each step of the blocking procedure, as will be explained in detail below.

We will see that after each RG step, the partition function will always have the shape \eqref{ZTNjf}, albeit with an updated numeric tensor on the coarse lattice. 
The RG steps are repeated until the tensor network has been reduced to a single tensor. The sum over the remaining indices of that tensor then yields the partition function $Z$.

\subsection{Coarsening the time direction}
\label{sec:contract1}

\subsubsection{Blocking adjacent tensors in the time direction}

As part of GHOTRG we now discuss an RG step in the \dir1-direction, which consists of (identical) contractions of all $V/2$ pairs of adjacent local tensors in that direction, 
\begin{align}
\cR^{\xx} \cG^{\xx} &\equiv
\sum_{\j_{x,1}}  \int_{c_{x,1}}
T^{(x)}_{\j_{x,-1}\j_{x,1}\j_{x,-2}\j_{x,2}}
T^{(x+\hat1)}_{\j_{x,1}\j_{x+\hat1,1}\j_{x+\hat1,-2}\j_{x+\hat1,2}}
K^{(x)}_{f_{x,-1}f_{x,1}f_{x,-2}f_{x,2}}
K^{(x+\hat1)}_{f_{x,1}f_{x+\hat1,1}f_{x+\hat1,-2}f_{x+\hat1,2}} 
,
\label{Tcontract1}
\end{align}
where $(x,x+\hat1)$ denotes the pair of sites which will eventually be fused in a new coarse-grained site, and $\cR^{\xx}$ and $\cG^{\xx}$ are the new numeric and Grassmann tensors, respectively, on the coarse lattice. 
The integral only represents an integration over the Grassmann variable $c_{x,1}$, which is defined on the link that connects the two sites. Note that the summation variable $j_{x,1}$ also appears in $f_{x,1}=f_{x,1}(j_{x,1})$.

We first consider the Grassmann part of this contraction, which is the product
\begin{align}
\cG^{\xx} &\equiv 
\int_{c_{x,1}} K^{(x)}_{f_{x,-1}f_{x,1}f_{x,-2}f_{x,2}} K^{(x+\hat1)}_{f_{x,1}f_{x+\hat1,1}f_{x+\hat1,-2}f_{x+\hat1,2}} .
\label{Kbarx}
\end{align}
The order of the two factors in the product is irrelevant as the Grassmann tensors can be considered to be commuting (as explained above), and we place them such that the Grassmann integration over the shared link can be directly performed,\begin{align}
\lefteqn{\cG^{\xx}}
\notag\\
&= \int_{c_{x,1}} \! (c_{x+\hat1,1})^{\f_{x+\hat1,1}} 
(c_{x+\hat1,2})^{\f_{x+\hat1,2}}
(dc_{x+\hat1,-2})^{\f_{x+\hat1,-2}}
(dc_{x,1})^{\f_{x,1}}
(c_{x,1})^{\f_{x,1}}
(c_{x,2})^{\f_{x,2}}
(dc_{x,-2})^{\f_{x,-2}}
(dc_{x,-1})^{\f_{x,-1}} 
\notag\\
&= (c_{x+\hat1,1})^{\f_{x+\hat1,1}} 
(c_{x+\hat1,2})^{\f_{x+\hat1,2}}
(dc_{x+\hat1,-2})^{\f_{x+\hat1,-2}}
(c_{x,2})^{\f_{x,2}}
(dc_{x,-2})^{\f_{x,-2}}
(dc_{x,-1})^{\f_{x,-1}} .
\label{Kx1a}
\end{align}
Note that $\cG^{\xx}$ does not depend on $f_{x,1}$ due to the integration formula \eqref{intaux}.
Therefore the sum over $j_{x,1}$ in \eqref{Tcontract1} actually only applies to the numeric tensors, such that
\begin{align}
\cR^{\xx} \equiv \cR^{\xx}_{\j_{x,-1}\j_{x+\hat1,1}(\j_{x,-2},\j_{x+\hat1,-2})(\j_{x,2},\j_{x+\hat1,2})} 
= 
\sum_{\j_{x,1}} 
T^{(x)}_{\j_{x,-1}\j_{x,1}\j_{x,-2}\j_{x,2}}
T^{(x+\hat1)}_{\j_{x,1}\j_{x+\hat1,1}\j_{x+\hat1,-2}\j_{x+\hat1,2}} .
\label{defcM}
\end{align}
In terms of the blocked tensors $\cR$ and $\cG$, the partition function is given by
\begin{align}
Z = \sum_{\boldsymbol j} \int \prod_{x\text{ odd}} \cR^{\xx} \cG^{\xx} 
\equiv \sum_{\boldsymbol j} \int \prod_{\X} \cR^{(\X)} \cG^{(\X)},
\label{ZMG}
\end{align}
where $\boldsymbol j$ now only contains all remaining indices, and $\X=\xx$ represents the sites on the coarse lattice.

Note that a tensor on the coarse lattice is connected to each neighbor in the contraction direction \dir1 by a single shared index $\j$ and to each neighbor in the perpendicular direction \dir2 by two such indices, which form ``fat indices" (in the following we therefore call the links in this direction ``fat links"). In \eqref{defcM} we denote the fat indices of $\cR$ by the pairs $(\j_{x,-2},\j_{x+\hat1,-2})$ and $(\j_{x,2},\j_{x+\hat1,2})$. In the following we want to apply the ideas of HOTRG to the blocked partition function \eqref{ZMG} and reduce the bond dimension of the fat indices from $D^2$ back to $D$, the bond dimension of the original indices.\footnote{Note that in the first coarsening step we actually truncate from $D_\text{initial}^2$ to $\min(D,D_\text{initial}^2)$, while in further steps the truncation will generically be from $D^2$ to $D$. For simplicity we will always refer to the generic case in the following.} Note that we cannot apply the HOSVD procedure as such to the coarse numeric tensor \eqref{defcM} because the coarse Grassmann tensor \eqref{Kx1a} depends on the same indices through $f(j)$.

By choosing the \dir1-direction as the first contraction direction, we always combine $T^{(x)}$ on a site with odd time coordinate with $T^{(x+\hat1)}$ on a site with even time coordinate, such that the contributions of the staggered phases are the same for all $\cR$ on the coarse lattice. 
Therefore, the new numeric tensor $\cR^{(\X)}$ is identical for all $\X$ on the coarse lattice. 
The new $\cG^{(\X)}$ can again be considered to be commuting in $Z$, as is explained in \ref{app:commuting}.

\subsubsection{Reducing the number of Grassmann variables in the space direction}

To reduce the number of Grassmann variables in the blocked Grassmann tensor, we will integrate out the Grassmann variables in the direction perpendicular to the contraction direction in \eqref{Kx1a}. However, the Grassmann variables and their corresponding differentials belong to tensors $\cG$ on different coarse sites in the partition function \eqref{ZMG}.
The differentials belonging to the fields $c_{x,2}$ and $c_{x+\hat1,2}$ in
$\cG^{(X)}$ can be found in $\cG^{(X+\hat 2)}$. Therefore, we want to reshuffle Grassmann differentials between all $\cG$ in the partition function \eqref{ZMG} to be able to integrate out the Grassmann variables in the \dir2-direction. 
To do so, the differentials
\begin{align}
(dc_{x+\hat1,-2})^{\f_{x+\hat1,-2}} 
\quad\text{and}\quad
(dc_{x,-2})^{\f_{x,-2}} 
\label{difs1}
\end{align}
will be moved from the coarse site $X$ to $X-\hat2$, and will be replaced by the differentials
\begin{align}
(dc_{x+\hat1,2})^{\f_{x+\hat1,2}}
\quad\text{and}\quad
(dc_{x,2})^{\f_{x,2}} ,
\label{difs2}
\end{align}
which are moved in from site $X+\hat 2$ to $X$. This applies to all $X$ on the coarse lattice.
This reshuffling of Grassmann differentials would however introduce non-local sign factors, and the partition function would no longer have the form of a tensor network.

To resolve this problem we define new auxiliary Grassmann variables $\cc$ on the fat links
of the coarse lattice by introducing a factor
\begin{align}
\left( \int d\cc_{X,-2} \cc_{X,-2} \right)^{\cf_{X,-2}} = 1
\label{fataux}
\end{align}
in every $\cG^{(\X)}$, with
\begin{align}
\cf_{X,-2} \equiv (\f_{x,-2} + \f_{x+\hat1,-2}) \!\!\!\mod 2 .
\label{fatmode}
\end{align}
Note that $\cf_{X,-2}$ is not an independent variable, but just an alias for the expression in \eqref{fatmode}, which we call the Grassmann parity of the fat index $(j_{x,-2},j_{x+\hat1,-2})$.

This definition guarantees that the sum $\f_{x,-2} + \f_{x+\hat1,-2} + \cf_{X,-2}$ is even, such that the product
\begin{align}
(\cc_{X,-2})^{\cf_{X,-2}}(dc_{x+\hat1,-2})^{\f_{x+\hat1,-2}}(dc_{x,-2})^{\f_{x,-2}} 
\label{comex}
\end{align}
is commuting.
 
After introducing \eqref{fataux} in \eqref{Kx1a} and reordering the differentials and fields, we find
\begin{align}
\cG^{(\X)} = 
\int_{\cc_{X,-2}}
&\sigma_{\f_{x,2}\f_{x+\hat1,-2}\f_{x+\hat1,2}}
(c_{x+\hat1,1})^{\f_{x+\hat1,1}}
(c_{x,2})^{\f_{x,2}}
(c_{x+\hat1,2})^{\f_{x+\hat1,2}}
\left( d\cc_{X,-2} \cc_{X,-2} \right)^{\cf_{X,-2}} 
\notag\\&\times
(dc_{x+\hat1,-2})^{\f_{x+\hat1,-2}}
(dc_{x,-2})^{\f_{x,-2}}
(dc_{x,-1})^{\f_{x,-1}} 
\label{cGfat}
\end{align}
with a sign factor
\begin{align}
\sigma_{\f_{x,2}\f_{x+\hat1,-2}\f_{x+\hat1,2}} &= (-1)^{\f_{x,2}(\f_{x+\hat1,-2}+\f_{x+\hat1,2})} .
\label{sbarx}
\end{align}

The partition function \eqref{ZMG} can now be written as
\begin{align}
Z = \sum_{\boldsymbol j} \int \prod_{\X} \cS^{(\X)} \cH^{(\X)} ,
\label{ZTH}
\end{align}
with a modified numeric tensor
\begin{align}
\cS^{(\X)} \equiv \cS^{(\X)}_{\j_{x,-1}\j_{x+\hat1,1}(\j_{x,-2},\j_{x+\hat1,-2})(\j_{x,2},\j_{x+\hat1,2})}
&= \sigma_{\f_{x,2}\f_{x+\hat1,-2}\f_{x+\hat1,2}} 
\cR^{(\X)}_{\j_{x,-1}\j_{x+\hat1,1}(\j_{x,-2},\j_{x+\hat1,-2})(\j_{x,2},\j_{x+\hat1,2})}
\label{defcT}
\end{align}
and a new Grassmann tensor
\begin{align}
\cH^{(\X)} &= 
(c_{x+\hat1,1})^{\f_{x+\hat1,1}}
(c_{x,2})^{\f_{x,2}}
(c_{x+\hat1,2})^{\f_{x+\hat1,2}}
\left( d\cc_{X,-2} \cc_{X,-2} \right)^{\cf_{X,-2}} 
(dc_{x+\hat1,-2})^{\f_{x+\hat1,-2}}
(dc_{x,-2})^{\f_{x,-2}}
(dc_{x,-1})^{\f_{x,-1}} .
\label{cH}
\end{align}
Note that the tensor $\cH$ has the same six indices as $\cG$ since $\cf$ is defined by \eqref{fatmode}. The integral over $\cc_{X,-2}$ in \eqref{cGfat} is now part of the integral in $Z$.

We are now able to move the commuting combination \eqref{comex} from the coarse site $X$ to $X-\hat2$, for all $X$,  without generating any sign factors. Hence, in $\cH^{(X)}$ this combination is replaced by the commuting expression
\begin{align}
(\cc_{X,2})^{\cf_{X,2}}
(dc_{x+\hat1,2})^{\f_{x+\hat1,2}}
(dc_{x,2})^{\f_{x,2}} ,
\label{difs4}
\end{align}
which is moved in from site $X+\hat 2$ to $X$.
This is done for all $X$ on the coarse lattice.\footnote{
When the shift moves variables over the lattice boundary, the boundary condition of $c$ needs to be applied. We choose the boundary conditions of the new auxiliary variables $\cc$ such that the combination \eqref{comex} does not generate sign factors when crossing the boundary. This is guaranteed when $\cc$ has the same boundary conditions as $c$ since $\tilde f_{X,-2}=(f_{x,-2}+f_{x+\hat1,-2})\mod2$.\label{fnbccc}}
The partition function \eqref{ZTH} can now be written as
\begin{align}
Z = \sum_{\boldsymbol j} \int \prod_{\X} \cS^{(\X)} \cK^{(\X)} ,
\label{ZTK}
\end{align}
with a new Grassmann tensor,
\begin{align}
\lefteqn{\cK^{(\X)}}
\notag\\ 
&= \int_{c_{x,2},c_{x+\hat1,2}}
\hspace{-2.5ex} 
(c_{x+\hat1,1})^{\f_{x+\hat1,1}}
(\cc_{X,2})^{\cf_{X,2}} 
(dc_{x+\hat1,2})^{\f_{x+\hat1,2}}
(dc_{x,2})^{\f_{x,2}}
(c_{x,2})^{\f_{x,2}}
(c_{x+\hat1,2})^{\f_{x+\hat1,2}}
(d\cc_{X,-2})^{\cf_{X,-2}}  
(dc_{x,-1})^{\f_{x,-1}}
\notag\\
&= 
(c_{x+\hat1,1})^{\f_{x+\hat1,1}}
(\cc_{X,2})^{\cf_{X,2}} 
(d\cc_{X,-2})^{\cf_{X,-2}}  
(dc_{x,-1})^{\f_{x,-1}} ,
\label{Kbarx3}
\end{align}
where we moved the commuting combination \eqref{difs4} to the appropriate position to perform the Grassmann integrations over $c_{x,2}$ and $c_{x+\hat1,2}$ without generating additional sign factors.

The new Grassmann tensors $\cK^{(\X)}$ can always be considered to be commuting, as the entries of the corresponding numeric tensors $\cS^{(\X)}$ are nonzero only when (see \ref{app:commuting})
\begin{align}
(\cf_{X,-2}+\cf_{X,2}+\f_{x,-1}+\f_{x+\hat1,1}) \!\!\!\mod 2 = 0 .
\label{comidmaintext}
\end{align}

\subsubsection{HOSVD of the numeric tensors}
\label{sec:hosvd}

In the following we will show how to apply an HOSVD approximation to reduce the dimension of the coarse-lattice numeric tensor $\cS$, by truncating its fat indices $(j_{x,-2},j_{x+\hat1,-2})$ and $(j_{x,2},j_{x+\hat1,2})$. As these indices also occur in $\cK$, it may seem as if this procedure cannot be applied. 
However, after the integration in \eqref{Kbarx3}, the new Grassmann tensor $\cK^{(\X)}$ only depends on $j_{x,-2}$ and $j_{x+\hat1,-2}$ through the sum of their Grassmann parities in $\cf_{X,-2}$, see \eqref{fatmode}. Similarly, it only depends on $j_{x,2}$ and $j_{x+\hat1,2}$ through the sum of their Grassmann parities $\cf_{X,2}$. 
Therefore, truncations of $\cS^{(\X)}$ are now possible if we separately truncate subspaces with even and odd Grassmann parities $\cf_{X,2}$ and $\cf_{X,-2}$.

\begin{figure}
\centering
\includegraphics{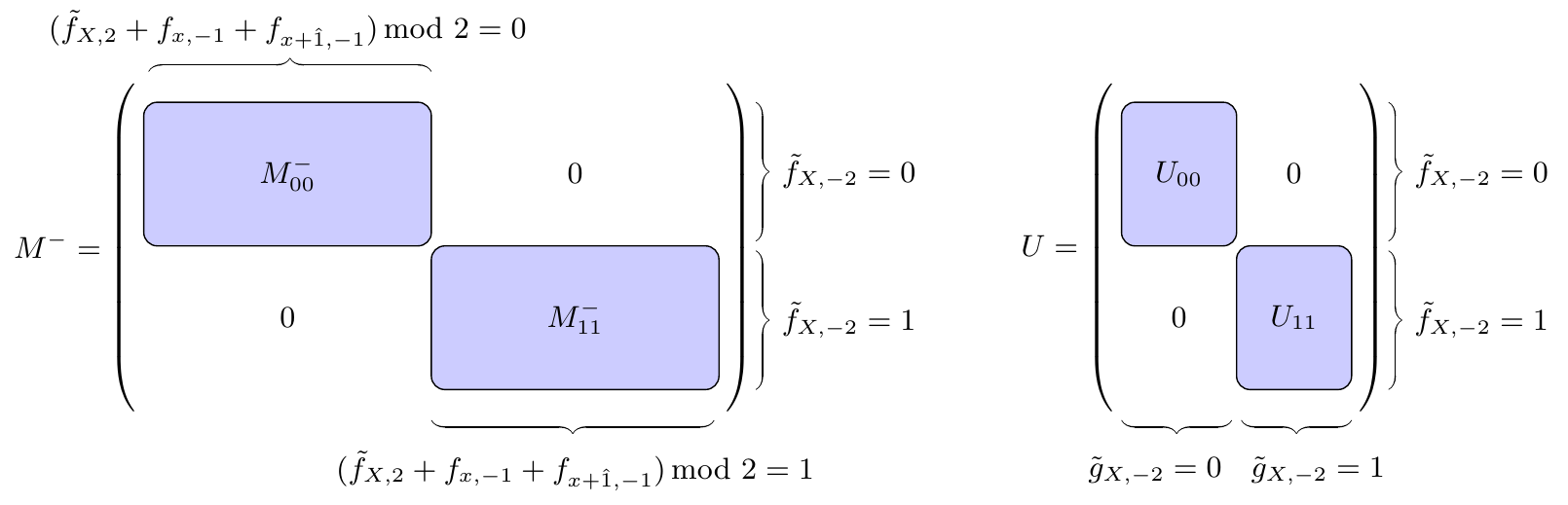}
\caption{The $D^2\times D^4$ dimensional matrization $M^-$ of $\cS$ has a row index $(\j_{x,-2},\j_{x+\hat1,-2})$ with corresponding Grassmann parity $\cf_{X,-2}$ and a column index $(\j_{x,-1},\j_{x+\hat1,1},(\j_{x,2},\j_{x+\hat1,2}))$ with corresponding Grassmann parity $(\cf_{X,2}+\f_{x,-1}+\f_{x+\hat1,1})\!\!\!\mod2$. The matrix $M^-$ is block diagonal with nonzero blocks corresponding to $\cf_{X,-2} = (\cf_{X,2}+\f_{x,-1}+\f_{x+\hat1,1})\!\!\!\mod2$.
The columns of the semi-orthogonal matrix $U$ are the $D$ leading left singular vectors $u^{(\cj_{X,-2})}$ of $M^-$. Therefore the $D^2\times D$ dimensional matrix $U$ is also block diagonal and its columns can be assigned a definite Grassmann parity $\cg_{X,-2}$.}
\label{fig:MU}
\end{figure}

Let us first analyze the HOSVD of the numerical tensor $\cS$.
The HOSVD procedure requires the computation of the left singular vectors of the matrizations $M$ of the coarse-lattice tensor $\cS$ of \eqref{defcT} with respect to its fat indices.
For a contraction in the \dir1-direction, the matrization with respect to the backward \dir2-direction yields the matrix
\begin{align}
M^{-}_{(\j_{x,-2},\j_{x+\hat1,-2})\,,\,(\j_{x,-1},\j_{x+\hat1,1},(\j_{x,2},\j_{x+\hat1,2}))}
&= \cS^{(\X)}_{\j_{x,-1}\j_{x+\hat1,1}(\j_{x,-2},\j_{x+\hat1,-2})(\j_{x,2},\j_{x+\hat1,2})} .
\end{align}
The matrix entries of $M^-$ are just a reordering of the tensor entries of $\cS$. From \eqref{comidmaintext} we see that the entries of $M^{-}$ are nonzero only when the Grassmann parities of its indices satisfy
\begin{align}
\cf_{X,-2} = (\cf_{X,2}+\f_{x,-1}+\f_{x+\hat1,1})\!\!\!\mod2 .
\label{parcons}
\end{align}
This means that the matrix $M^{-}$ is block diagonal\footnote{The matrix $M^{-}$ can be brought in block diagonal form by permutations of basis vectors.} with nonzero blocks corresponding to $\cf_{X,-2}=(\cf_{X,2}+\f_{x,-1}+\f_{x+\hat1,1})\mod2=0$ or $\cf_{X,-2}=(\cf_{X,2}+\f_{x,-1}+\f_{x+\hat1,1})\!\!\!\mod2=1$, see Fig.\ \ref{fig:MU} (left). Therefore the left singular vectors $u^{(\cj_{X,-2})}_{(\j_{x,-2},\j_{x+\hat1,-2})}$ of $M^{-}$, for which we introduce the label by $\cj_{X,-2}$, can be assigned definite Grassmann parities $\cg_{X,-2}$ since the nonzero entries of a single singular vector all have the same Grassmann parity $\cf_{X,-2}=0$ or $\cf_{X,-2}=1$, see Fig.\ \ref{fig:MU} (right).\footnote{Note that in the case of degenerate singular values, one can always choose a basis consisting of vectors with definite Grassmann parities.} This establishes a map from $\cj_{X,-2}$ to the Grassmann parity $\cg_{X,-2}$.

The same reasoning applies for the matrization $M^{+}_{(\j_{x,2},\j_{x+\hat1,2})\,,\,(\j_{x,-1},\j_{x+\hat1,1},(\j_{x,-2},\j_{x+\hat1,-2}))}$ in the forward \dir2-direction, with nonzero blocks for $\cf_{X,2}=(\cf_{X,-2}+\f_{x,-1}+\f_{x+\hat1,1})\!\!\!\mod2$. Furthermore, the relation between $j_{x,-2}$ and $\f_{x,-2}$ is the same as that between $j_{x,2}$ and $\f_{x,2}$ for all $x$.\footnote{This is so by construction for the initial local tensor, and remains so throughout the blocking procedure by applying the same truncation matrices $U$ to the forward and backward directions.} Therefore $M^{-}$ and $M^{+}$ have the same block structure.

Nevertheless, the matrices $M^{-}$ and $M^{+}$ generically have different singular values and singular vectors. However, for HOTRG we have to project the vector spaces of dimensions $D^2$ belonging to the backward and forward directions on the same $D$-dimensional subspace. Hence, a common $D^2\times  D$ semi-orthogonal truncation matrix $U$ has to be constructed from $M^{-}$ and $M^{+}$. The standard approach \cite{Xie_2012} consists of constructing $U$ with the $D$ leading left singular vectors of $M^-$ or $M^+$, depending on which one yields the smallest truncation error, i.e., the largest value for the sum of their $D$ largest singular values. However, we have developed a so-called SuperQ method \cite{Bloch:2022}, which reduces the combined local approximation error, defined below in \eqref{apperror}, by constructing $U$ with the leading left singular vectors of the extended matrix $M=(M^- \,\, M^+)$. As the Gram matrices $Q^-=M^{-}(M^{-})^T$ and $Q^+=M^{+}(M^{+})^T$ have identical Grassmann parity block structures, so will $Q=M M^T=Q^-+Q^+$. In all cases the truncation matrix $U$ is populated by $D$ orthonormal column vectors with column indices $\cj$, which can always be assigned definite Grassmann parities $\cg$ (as explained in detail above for $M^-$).

Therefore, in $\cS$ of \eqref{defcT} we can truncate the fat indices $(\j_{x,-2},\j_{x+\hat1,-2})$ and $(\j_{x,2},\j_{x+\hat1,2})$ with dimension $D^2$ to new thin indices $\cj_{\X,-2}$ and $\cj_{\X,2}$ of dimension $D$ with Grassmann parities $\cg_{\X,-2}\equiv\cg_{\X,-2}(\cj_{\X,-2})$ and $\cg_{\X,2}\equiv\cg_{\X,2}(\cj_{\X,2})$ and construct a new tensor
\begin{align}
\cT^{(\X)}_{\j_{x,-1}\j_{x+\hat1,1}\cj_{\X,-2}\cj_{\X,2}}
&= \sum_{\substack{\j_{x,-2},\j_{x+\hat1,-2},\\\j_{x,2},\j_{x+\hat1,2}}} U_{(\j_{x,-2},\j_{x+\hat1,-2})\cj_{\X,-2}} U_{(\j_{x,2},\j_{x+\hat1,2})\cj_{\X,2}} 
\cS^{(\X)}_{\j_{x,-1}\j_{x+\hat1,1}(\j_{x,-2},\j_{x+\hat1,-2})(\j_{x,2},\j_{x+\hat1,2})} .
\label{truncT}
\end{align}
The error of the combined truncations of the fat indices can be quantified by the Frobenius norm 
\begin{align}
&\left\| \sum_{\cj_{\X,-2},\cj_{\X,2}} U_{(\j_{x,-2},\j_{x+\hat1,-2})\cj_{\X,-2}} U_{(\j_{x,2},\j_{x+\hat1,2})\cj_{\X,2}} \cT^{(\X)}_{\j_{x,-1}\j_{x+\hat1,1}\cj_{\X,-2}\cj_{\X,2}} 
 -
 \cS^{(\X)}_{\j_{x,-1}\j_{x+\hat1,1}(\j_{x,-2},\j_{x+\hat1,-2})(\j_{x,2},\j_{x+\hat1,2})}
\right\| .
\label{apperror}
\end{align}
Note that in \eqref{truncT} the same semi-orthogonal matrix $U$ is used for backward and forward directions, such that the tensor network on the coarse lattice can be written in terms of $\cT$, with bond dimensions $D$ on all links.\footnote{If we would use different matrices $U^-$ and $U^+$ for the backward and forward directions, the tensor network would not only consist of $\cT$ but would also explicitly depend on $U^\pm$.}

\subsubsection{Applying the truncation matrices}
\label{sec:applyU}

As the indices of the numeric tensor also appear in the Grassmann tensor, we need to determine the effect of applying the truncation matrix $U$ to the product of the two tensors.
We first consider the application of $U$ to truncate the fat index for the backward direction,
\begin{align}
\sum_{\j_{x,-2},\j_{x+\hat1,-2}} U_{(\j_{x,-2},\j_{x+\hat1,-2})\cj_{\X,-2}} 
\cS^{(\X)}_{\j_{x,-1}\j_{x+\hat1,1}(\j_{x,-2},\j_{x+\hat1,-2})(\j_{x,2},\j_{x+\hat1,2})}
\cK^{(\X)}_{\f_{x,-1}\f_{x+\hat1,1}\cf_{\X,-2}\cf_{\X,2}} .
\label{UTK}
\end{align}
Due to the block-diagonal nature of $U$, we observe that for those $\cj_{\X,-2}$ which have Grassmann parity $\cg_{\X,-2}=0$, only fat indices $(\j_{x,-2},\j_{x+\hat1,-2})$ with Grassmann parity $\cf_{\X,-2}=0$ contribute to the sum. Similarly, for $\cj_{\X,-2}$ with $\cg_{\X,-2}=1$, only fat indices $(\j_{x,-2},\j_{x+\hat1,-2})$ with $\cf_{\X,-2}=1$ result in nonzero contributions. Therefore we can replace the index $\cf_{\X,-2}$ of $\cK$ with $\cg_{\X,-2}$, such that \eqref{UTK} becomes 
\begin{align}
\cK^{(\X)}_{\f_{x,-1}\f_{x+\hat1,1}\cg_{\X,-2}\cf_{\X,2}} \sum_{\j_{x,-2},\j_{x+\hat1,-2}} U_{(\j_{x,-2},\j_{x+\hat1,-2})\cj_{\X,-2}} 
\cS^{(\X)}_{\j_{x,-1}\j_{x+\hat1,1}(\j_{x,-2},\j_{x+\hat1,-2})(\j_{x,2},\j_{x+\hat1,2})}
.
\end{align}
Similarly, for the truncation in the forward direction, $\cf_{\X,2}$ can be replaced by $\cg_{\X,2}$ in $\cK$. After these replacements the truncation matrix $U$ only acts on the numeric tensor $\cS$, which leads to the truncated numeric tensor $\cT$ of \eqref{truncT}.
The Grassmann parities $\cg$ of the indices $\cj$ of $\cT$ become the new coarse site Grassmann indices in the Grassmann tensor $\cK$.

\begin{figure}
\begin{center}
\includegraphics{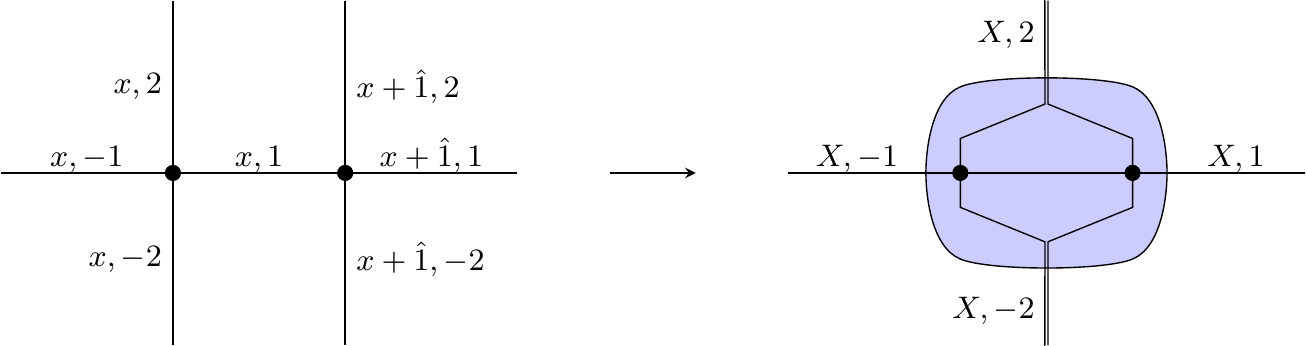}
\caption{Illustration of a contraction in the \dir1-direction and the corresponding coarsening.}
\label{Fig:coarsening}
\end{center}
\end{figure}

As explained above, the coarse tensor $\cS$ is identical for all sites $\X$ on the coarsened lattice. Therefore the truncation procedure is identical for all these sites.
The links $(x,-1)$ and $(x+\hat1,1)$ on the original lattice become $(\X,-1)$ and $(\X,1)$ on the coarse lattice, see Fig.~\ref{Fig:coarsening}, such that the new coarse local tensor $\cT$ has the following entries for the coarse site $\X = (x,x+\hat1)$,
\begin{align}
\cT^{(\X)}_{\j_{\X,-1}\j_{\X,1}\clf_{\X,-2}\clf_{\X,2}} ,
\end{align} 
where $\cj$ are the new indices introduced in the truncation procedure \eqref{truncT}. 
Using the same change of notation, the Grassmann tensor \eqref{Kbarx3} on the coarse lattice becomes
\begin{align}
\cK^{(\X)}_{\f_{\X,-1}\f_{\X,1}\cg_{\X,-2} \cg_{\X,2}}  =
(c_{\X,1})^{\f_{\X,1}}
(\cc_{\X,2})^{\cg_{\X,2}}  
(d\cc_{\X,-2})^{\cg_{\X,-2}} 
(dc_{\X,-1})^{\f_{\X,-1}} 
\end{align}
with $\f_{\X,\pm 1}\equiv\f_{\X,\pm 1}(\j_{\X,\pm 1})$ and $\cg_{\X,\pm 2}\equiv\cg_{\X,\pm 2}(\cj_{\X,\pm 2})$.
As is explained in \ref{app:commuting}, the tensors $\cK$ can still be considered to be commuting.
The partition function on the coarse lattice then reads
\begin{align}
  Z = \sum_{\boldsymbol j,\boldsymbol \cj} \int \prod_{\X=1}^{\cV}  \cT^{(\X)}_{\j_{\X,-1}\j_{\X,1}\clf_{\X,-2}\clf_{\X,2}}\ \cK^{(\X)}_{\f_{\X,-1}\f_{\X,1}\cg_{\X,-2} \cg_{\X,2}} ,
 \label{finalZcoarse}
\end{align}
where $\cV=V/2$.
We can now rename  
\begin{align}
  \cT \to T,\quad
  \cK \to K,\quad
  \cc_{\X,2} \to c_{\X,2},\quad
  \cj_{\X,2} \to j_{\X,2},\quad
  \cg_{\X,2} \to f_{\X,2}\quad
\label{renaming}
\end{align}
and finally
\begin{align}
\quad X \to x,\quad \cV \to V.
\end{align}
After this change of notation, the partition function \eqref{finalZcoarse} has the exact same form as the original $Z$ in \eqref{ZTNjf}, albeit now on the coarsened lattice of half the volume.
This means that the coarsening procedure detailed above is self-reproducing.
Below we will show that the same holds for contractions in the \dir2-direction. Therefore, the blocking steps in either direction can be repeated iteratively using the exact same manipulations until the complete lattice has been reduced to a single site. 

Note that all sign factors generated by Grassmann manipulations only depend on local indices, i.e., indices connected to the sites being contracted. This property, which allows us to absorb the sign factor in the new numeric tensor on the coarse lattice, is crucial for the application of the iterative renormalization group procedure.

The implementation of the GHOTRG algorithm is described in \ref{app:implementation}. To improve its efficiency, the truncated coarse-lattice tensor $\cT$ is constructed without explicitly computing the full coarse-lattice tensors $\cR$ of \eqref{defcM} nor $\cS$ of \eqref{defcT}.

\subsection{Coarsening the space direction}
\label{sec:contract2}

The contraction procedure detailed in Sec.~\ref{sec:contract1} for a contraction in the \dir1-direction can also be applied to perform a contraction in the \dir2-direction.%
\footnote{Note that after a first contraction in the \dir1-direction, the local numeric tensors are identical for all sites, see Sec.~\ref{sec:contract1}.}
To this end, we first reorder the Grassmann fields in \eqref{Kx} such that the directions are exchanged,
\begin{align}
K^{(x)}_{f_{x,-1}f_{x,1}f_{x,-2}f_{x,2}} &= 
(c_{x,1})^{\f_{x,1}}
(c_{x,2})^{\f_{x,2}}  
(dc_{x,-2})^{\f_{x,-2}} 
(dc_{x,-1})^{\f_{x,-1}}
\notag\\
&=
\hat\sigma_{\f_{x,-1}\f_{x,1}\f_{x,-2}\f_{x,2}} 
(c_{x,2})^{\f_{x,2}}  
(c_{x,1})^{\f_{x,1}}
(dc_{x,-1})^{\f_{x,-1}}
(dc_{x,-2})^{\f_{x,-2}} 
\label{Kx2}
\end{align}
with sign factor
\begin{align}
\hat\sigma_{\f_{x,-1}\f_{x,1}\f_{x,-2}\f_{x,2}} = (-1)^{\f_{x,1}\f_{x,2}+\f_{x,-1}\f_{x,-2}} .
\label{hatsigma}
\end{align}

After defining a new Grassmann tensor
\begin{align}
\widehat K^{(x)}_{f_{x,-2}f_{x,2}f_{x,-1}f_{x,1}}
 =
(c_{x,2})^{\f_{x,2}}  
(c_{x,1})^{\f_{x,1}}
(dc_{x,-1})^{\f_{x,-1}}
(dc_{x,-2})^{\f_{x,-2}} 
\end{align}
and a new numeric tensor
\begin{align}
\widehat T^{(x)}_{\j_{x,-2}\j_{x,2}\j_{x,-1}\j_{x,1}}
=
\hat\sigma_{\f_{x,-1}\f_{x,1}\f_{x,-2}\f_{x,2}} \, T^{(x)}_{\j_{x,-1}\j_{x,1}\j_{x,-2}\j_{x,2}} ,
\end{align}
the partition function \eqref{ZTNjf} can be rewritten as 
\begin{align}
Z = \sum_{\boldsymbol j} \int \prod_x
\widehat T^{(x)}_{\j_{x,-2}\j_{x,2}\j_{x,-1}\j_{x,1}}
\widehat K^{(x)}_{f_{x,-2}f_{x,2}f_{x,-1}f_{x,1}} .
\label{ZTNjf2}
\end{align}
We observe that the partition function \eqref{ZTNjf2} has a structure identical to \eqref{ZTNjf}, up to an exchange of the directions $1\leftrightarrow 2$ and a renaming of $T\to\widehat T$ and $K\to\widehat K$. 

From here on, everything derived in Sec.~\ref{sec:contract1} for a contraction in the \dir1-direction can be applied to a contraction in the \dir2-direction by just exchanging $1\leftrightarrow 2$ everywhere.
This means that we again integrate out the Grassmann field along the contracted link, introduce new Grassmann variables $\cc_{\X,1}$ on the fat links perpendicular to the contraction direction, move the differentials one site backward in the \dir1-direction, and integrate out the old Grassmann variables in that direction.
 The coarse numeric tensor is truncated using HOSVD which yields the partition function \eqref{finalZcoarse} with directions \dir1 and \dir2 exchanged. Similarly to \eqref{renaming} we rename
\begin{align}
  \cT \to \widehat T,\quad
  \cK \to \widehat K,\quad
  \cc_{\X,1} \to c_{\X,1},\quad
  \cj_{\X,1} \to j_{\X,1},\quad
  \cg_{\X,1} \to f_{\X,1}\quad
\label{renaming2}
\end{align}
and again $X \to x$ and $\cV \to V$. Then, the partition function on the coarse lattice is identical to \eqref{ZTNjf2}, albeit on a lattice of half the volume.

We now convert the Grassmann tensor back to its canonical form \eqref{Kx}, such that further blockings in either direction can be applied, using the procedures detailed in Secs.~\ref{sec:contract1} and \ref{sec:contract2}.
After a contraction in the \dir2-direction, the coarse Grassmann tensor reads
\begin{align}
\widehat K^{(x)}_{f_{x,-2}f_{x,2}f_{x,-1}f_{x,1}}
 &= 
(c_{x,2})^{\f_{x,2}}  
(c_{x,1})^{\f_{x,1}}
(dc_{x,-1})^{\f_{x,-1}}
(dc_{x,-2})^{\f_{x,-2}} 
\notag\\
&=
\hat\sigma_{\f_{x,-1}\f_{x,1}\f_{x,-2}\f_{x,2}} 
(c_{x,1})^{\f_{x,1}}
(c_{x,2})^{\f_{x,2}}  
(dc_{x,-2})^{\f_{x,-2}} 
(dc_{x,-1})^{\f_{x,-1}}
\end{align}
with sign factor $\hat\sigma$ given in \eqref{hatsigma}.
After defining a new Grassmann tensor
\begin{align}
K^{(x)}_{f_{x,-1}f_{x,1}f_{x,-2}f_{x,2}}
 =
(c_{x,1})^{\f_{x,1}}
(c_{x,2})^{\f_{x,2}}  
(dc_{x,-2})^{\f_{x,-2}} 
(dc_{x,-1})^{\f_{x,-1}} ,
\end{align}
which is in the canonical form, and a new numeric tensor
\begin{align}
 T^{(x)}_{\j_{x,-1}\j_{x,1}\j_{x,-2}\j_{x,2}}
=
\hat\sigma_{\f_{x,-1}\f_{x,1}\f_{x,-2}\f_{x,2}} 
\widehat T^{(x)}_{\j_{x,-2}\j_{x,2}\j_{x,-1}\j_{x,1}}
\, ,
\end{align}
the partition function on the coarse lattice again has its original form \eqref{ZTNjf}.

The use of a canonical order for the variables in the Grassmann tensor conveniently allows for a flexible order of contraction directions.

\subsection{Blocking the complete lattice and applying the boundary conditions}
\label{sec:bc}

We can now repeat contractions in both directions according to the procedures described in Secs.~\ref{sec:contract1} and \ref{sec:contract2}, until the complete lattice has been reduced to a single site. On the remaining site, the backward and forward links are identical such that $j_{x,\nu}=j_{x,-\nu}$ (and correspondingly $f_{x,\nu}=f_{x,-\nu}$), and the partition function \eqref{ZTNjf} reduces to
\begin{align}
Z = \sum_{\j_{x,1},\j_{x,2}} \int T_{\j_{x,1}\j_{x,1}\j_{x,2}\j_{x,2}} K_{\f_{x,1}\f_{x,1}\f_{x,2}\f_{x,2}} ,
\label{finalTK}
\end{align}
with
\begin{align}
K_{\f_{x,1}\f_{x,1}\f_{x,2}\f_{x,2}} = 
(c_{x,1})^{\f_{x,1}}
(c_{x,2})^{\f_{x,2}} 
(dc_{x,-2})^{\f_{x,2}} 
(dc_{x,-1})^{\f_{x,1}}
\label{Ksingle}
.
\end{align}
The boundary conditions on the Grassmann variables are easily applied in our version of the GHOTRG procedure. As we have shown in the sections above, the boundary conditions are automatically transferred to the coarse site Grassmann variables $c_{x,\nu}$ at each coarsening step.
Therefore, the antiperiodic boundary conditions in time are implemented by imposing $c_{x,-1}=-c_{x,1}$ in \eqref{Ksingle}, while the periodic boundary conditions in space are given by $c_{x,-2}=c_{x,2}$. This results in
\begin{align}
\int K_{\f_{x,1}\f_{x,1}\f_{x,2}\f_{x,2}} = 
\int (c_{x,1})^{\f_{x,1}}
(c_{x,2})^{\f_{x,2}} 
(dc_{x,2})^{\f_{x,2}} 
(-dc_{x,1})^{\f_{x,1}}
= (-1)^{f_{x,2}}  ,
\end{align}
and the partition function is thus given by
\begin{align}
Z = \sum_{\j_{x,1},\j_{x,2}} (-1)^{f_{x,2}} T_{\j_{x,1}\j_{x,1}\j_{x,2}\j_{x,2}} .
\label{finalZwithBC}
\end{align}
In analogy to matrices, the sums in \eqref{finalZwithBC} are often referred to as tensor traces in the corresponding directions.

\subsection{ASAP-tracing}

As an alternative to the procedure described in Sec.\ \ref{sec:bc}, we can also apply ``ASAP-tracing".
As soon as the lattice has been reduced to a single one-dimensional slice, the tensor can be traced out in the perpendicular direction. 
This slightly improves the accuracy of the GHOTRG method since it avoids unnecessary truncations that would otherwise arise in the further coarsening steps of the remaining direction. When performing this ASAP-tracing, one first integrates out the Grassmann variables in the perpendicular direction taking into account the boundary conditions. We start from the Grassmann tensor in its canonical form,
\begin{align}
K_{\f_{x,-1}\f_{x,1}\f_{x,-2}\f_{x,2}} = 
(c_{x,1})^{\f_{x,1}}
(c_{x,2})^{\f_{x,2}} 
(dc_{x,-2})^{\f_{x,-2}} 
(dc_{x,-1})^{\f_{x,-1}} ,
\label{Kcanon}
\end{align}
and consider a tracing in either the \dir1- or \dir2-direction below.

\subsubsection{Tracing the time direction}

For a slice in the \dir2-direction, we reorder the Grassmann variables in \eqref{Kcanon} to integrate out the variables in the \dir1-direction with $f_{x,1}=f_{x,-1}$ and apply the antiperiodic boundary conditions in time by setting $c_{x,-1}=-c_{x,1}$. This leads to
\begin{align}
\int_{c_{x,1}} K_{\f_{x,1}\f_{x,1}\f_{x,-2}\f_{x,2}} &= 
\int_{c_{x,1}} (c_{x,1})^{\f_{x,1}}
(c_{x,2})^{\f_{x,2}} 
(dc_{x,-2})^{\f_{x,-2}} 
(-dc_{x,1})^{\f_{x,1}}
\notag\\
&=  (-1)^{\f_{x,1}(\f_{x,2}+\f_{x,-2})} (c_{x,2})^{\f_{x,2}} (dc_{x,-2})^{\f_{x,-2}} .
\end{align}
We now define
\begin{align}
K^\text{(1d)}_{f_{x,-2}f_{x,2}} =  (c_{x,2})^{\f_{x,2}} (dc_{x,-2})^{\f_{x,-2}}
\label{K1d}
\end{align}
and
\begin{align}
T^\text{(1d)}_{\j_{x,-2}\j_{x,2}} = \sum_{\j_{x,1}} (-1)^{\f_{x,1}(\f_{x,2}+\f_{x,-2})} T_{\j_{x,1}\j_{x,1}\j_{x,-2}\j_{x,2}}
\end{align}
such that the partition function is
\begin{align}
Z = \sum_{\boldsymbol j} T^\text{(1d)}_{\j_{x,-2}\j_{x,2}} K^\text{(1d)}_{f_{x,-2}f_{x,2}}.
\label{Z1d}
\end{align}
Only the entries of the matrix $T^\text{(1d)}$ with even $\f_{x,-2}+\f_{x,2}$ are nonzero, see \ref{app:commuting}, such that $K^\text{(1d)}$ can be considered to be commuting.

In the subsequent spatial tensor contractions, the Grassmann tensor is given by
\begin{align}
\cK^{(1d)}_{f_{\X,-2}f_{\X,2}} 
&= \int_{c_{x,2}}  K^\text{(1d)}_{f_{x,2}f_{x+\hat2,2}} K^\text{(1d)}_{f_{x,-2}f_{x,2}}
= \int_{c_{x,2}} (c_{x_+\hat2,2})^{\f_{x+\hat2,2}} (dc_{x,2})^{\f_{x,2}}  (c_{x,2})^{\f_{x,2}} (dc_{x,-2})^{\f_{x,-2}} 
\notag\\
&= (c_{x+\hat2,2})^{\f_{x+\hat2,2}}(dc_{x,-2})^{\f_{x,-2}} 
\equiv (c_{\X,2})^{\f_{\X,2}}(dc_{\X,-2})^{\f_{\X,-2}} ,
\end{align}
and after taking $\X \to x$ and $\cK\to K$ the Grassmann tensor is identical to \eqref{K1d}, albeit with $x$ now on the coarse lattice. This means that the Grassmann tensor is self-reproducing in the one-dimensional coarsening procedure. 
The contraction of the two adjacent numeric tensors then yields (the Grassmann tensor no longer depends on $f_{x,2}$),
\begin{align}
\cT^{(1d)}_{\j_{\X,-2}\j_{\X,2}} &= \sum_{\j_{x,2}} T^\text{(1d)}_{\j_{x,-2}\j_{x,2}} T^\text{(1d)}_{\j_{x,2}\j_{x+\hat2,2}} ,
\end{align}
which corresponds to a matrix multiplication. When taking $X\to x$ and $\cT\to T$, the partition function again looks like \eqref{Z1d}, albeit on the coarsened lattice.

After contracting the remaining sites in the \dir2-direction until only one site is left, the final trace with periodic boundary conditions in the spatial direction yields 
\begin{align}
Z 
= \sum_{\j_{x,2}} T^\text{(1d)}_{\j_{x,2}\j_{x,2}} \int_{c_{x,2}} K^\text{(1d)}_{f_{x,2}f_{x,2}} = 
\sum_{\j_{x,2}} T^\text{(1d)}_{\j_{x,2}\j_{x,2}} \int_{c_{x,2}} (c_{x,2})^{\f_{x,2}}(dc_{x,2})^{\f_{x,2}}
= \sum_{\j_{x,2}} (-1)^{f_{x,2}} T^\text{(1d)}_{\j_{x,2}\j_{x,2}} .
\end{align}

\subsubsection{Tracing the space direction}

We now consider a slice in the \dir1-direction. Since we use periodic boundary conditions in space, we have $c_{x,-2}=c_{x,2}$. After reordering the Grassmann variables to integrate out the variables in the \dir2-direction, with $f_{x,-2}=f_{x,2}$, we obtain
\begin{align}
\int_{c_{x,2}} K_{\f_{x,-1}\f_{x,1}\f_{x,2}\f_{x,2}} = 
\int_{c_{x,2}}
(c_{x,1})^{\f_{x,1}}
(c_{x,2})^{\f_{x,2}} 
(dc_{x,2})^{\f_{x,2}} 
(dc_{x,-1})^{\f_{x,-1}}
= (-1)^{\f_{x,2}} (c_{x,1})^{\f_{x,1}} (dc_{x,-1})^{\f_{x,-1}} .
\end{align}
In this case we define the Grassmann and numeric tensors on the remaining one-dimensional lattice as
\begin{align}
K^\text{(1d)}_{f_{x,-1}f_{x,1}} = (c_{x,1})^{\f_{x,1}} (dc_{x,-1})^{\f_{x,-1}}
\label{K1dx}
\end{align}
and
\begin{align}
T^\text{(1d)}_{\j_{x,-1}\j_{x,1}} = \sum_{\j_{x,2}} (-1)^{\f_{x,2}} T_{\j_{x,-1}\j_{x,1}\j_{x,2}\j_{x,2}}
\end{align}
such that the partition function can be written as
\begin{align}
Z = \sum_{\boldsymbol j} T^\text{(1d)}_{\j_{x,-1}\j_{x,1}} K^\text{(1d)}_{f_{x,-1}f_{x,1}}.
\label{Z1dx}
\end{align}
When blocking sites in the remaining \dir1-direction, the product of Grassmann tensors is self-reproducing, as
\begin{align}
\cK^\text{(1d)}_{f_{X,-1}f_{X,1}} 
&= \int_{c_{x,1}} K^\text{(1d)}_{f_{x,1}f_{x+\hat1,1}} K^\text{(1d)}_{f_{x,-1}f_{x,1}} 
= \int_{c_{x,1}} (c_{x_+\hat1,1})^{\f_{x+\hat1,1}} (dc_{x,1})^{\f_{x,1}}  (c_{x,1})^{\f_{x,1}} (dc_{x,-1})^{\f_{x,-1}} 
\notag\\
&= (c_{x_+\hat1,1})^{\f_{x+\hat1,1}}(dc_{x,-1})^{\f_{x,-1}} 
\equiv (c_{\X,1})^{\f_{\X,1}}(dc_{\X,-1})^{\f_{\X,-1}} ,
\end{align}
which after taking $\X\to x$ and $\cK \to K$ again yields the structure of \eqref{K1dx}, albeit with $x$ on the coarsened lattice. The contraction of the two adjacent numeric tensors then yields (the Grassmann tensor no longer depends on $f_{x,1}$),
\begin{align}
\cT^{(1d)}_{\j_{\X,-1}\j_{\X,1}} &= \sum_{\j_{x,1}} T^\text{(1d)}_{\j_{x,-1}\j_{x,1}} T^\text{(1d)}_{\j_{x,1}\j_{x+\hat1,1}} ,
\end{align}
which is a matrix multiplication. When taking $X\to x$ and $\cT\to T$, the partition function again has the form of \eqref{Z1dx}, albeit on the coarsened lattice.

After contracting the remaining sites in the \dir1-direction, the final trace with antiperiodic boundary conditions in the time direction yields 
\begin{align}
Z 
= \sum_{\j_{x,1}} T^\text{(1d)}_{\j_{x,1}\j_{x,1}} \int_{c_{x,1}} K^\text{(1d)}_{f_{x,1}f_{x,1}} = 
\sum_{\j_{x,1}} T^\text{(1d)}_{\j_{x,1}\j_{x,1}} \int_{c_{x,1}} (-c_{x,1})^{\f_{x,1}}(dc_{x,1})^{\f_{x,1}}
= \sum_{\j_{x,1}} T^\text{(1d)}_{\j_{x,1}\j_{x,1}} .
\end{align}

\section{Results}
\label{sec:results}

In this section we report about the application of our GHOTRG method for two-dimensional strong-coupling QCD with staggered quarks, where we set the anisotropy factor $\gamma=1$.

We first consider a baryon-only version of the model to validate the Grassmann blocking without being affected by possibly large mesonic contributions. For small lattices, the numerical results are verified with exact analytic computations. 

Next we report about the application of the GHOTRG method to the full strong-coupling meson-baryon system and again compare with exact results for small lattices. Furthermore, we investigate the  convergence of $\log(Z)/V$ with the bond dimension $D$.
Besides the partition function itself, we also compute the chiral condensate
\begin{align}
\braket{\bar\psi\psi} = \frac1V \frac{\partial \logZ}{\partial m}
\label{cc}
\end{align}
and the quark number density
\begin{align}
\rho = \frac1V \frac{\partial \logZ}{\partial\mu} .
\label{rho}
\end{align}
For large volumes, we study the behavior of the chiral condensate as a function of the mass and the volume at zero chemical potential, in order to investigate the chiral symmetry of the model. Finally we present results for the quark number density and the chiral condensate as a function of the chemical potential and obtain some evidence for a first-order phase transition.

We implemented our version of the GHOTRG procedure as an extension to our already existing C++ HOTRG library. Some specifics of our implementation are described in \ref{app:implementation}.

\subsection{Computing observables with stabilized finite differences}

In tensor studies, observables are often computed using finite differences of $\logZ$ or using an impurity method. To overcome the drawbacks of both methods, we developed a stabilized finite-difference (SFD) method \cite{Bloch:2021mjw}.
To motivate the method, it is useful to describe the problem encountered with the traditional finite-difference computations in tensor methods.  Numerical finite differences only work properly for functions that are sufficiently smooth. However, the very nature of tensor-network methods is that discrete truncations are applied during the blocking procedure, and these truncations very easily break the required smoothness property of $\logZ$ with varying parameter values. 
The problem occurs when the computed $\logZ$ jumps between close-by parameter values required for the evaluation of finite differences. In tensor methods, such jumps are typically caused by (almost-)degenerate singular values and/or level crossings of singular values, which lead to discontinuous changes of the vector subspaces used to truncate the coarse-lattice tensors.
This problem can in principle only be resolved by taking the bond dimension $D$ so large that the systematical error on $\logZ$ is much smaller than the difference between the exact values of $\logZ$ for two different parameter values. If such a bond dimension cannot be achieved, as is often the case, the computed finite differences will have large errors.
 
A solution to this problem, which we developed with the SFD method, is to modify the HOSVD truncations in order to improve the smoothness properties of the computed $\logZ$, required for the application of the finite-difference method.
The stabilization uses a heuristic approach that operates on the singular vectors of HOSVD to maximize the overlap between the truncated vector spaces constructed for adjacent parameters values. This is achieved by considering almost-degenerate singular values for both parameter values, and introducing separate basis changes in the respective subspaces. 
The method uses the fact that small variations of the parameter values generically lead to small rotations of these subspaces and allows for the use of very small step sizes in the finite-difference formula. 

Note that observables can also be computed using the impurity method. Although this method yields smoother data (which does not necessarily mean more accurate) than the non-stabilized finite-difference method, it has its own systematic error because the same singular vectors are used to truncate the pure and impure tensors. We therefore use the SFD method as method of choice to compute observables. The SFD method was also used successfully to stabilize second-order finite differences in the computation of susceptibilities, e.g., the specific heat of the three-dimensional O(2) model \cite{Bloch:2021mjw}.

\subsection{Baryon-only partition function}
\label{sec:baryon-only}

As a first validation of our GHOTRG method for strong-coupling QCD, we discard all mesonic contributions in \eqref{ZTNjf}, i.e., we replace $\delta_{x\in{\cal M}} \to 0$ in \eqref{Tm}.
Then, the resulting partition function is independent of the mass and all sites of contributing configurations are baryonic. 

\begin{figure}
\centerline{
\includegraphics[width=0.499\hsize]{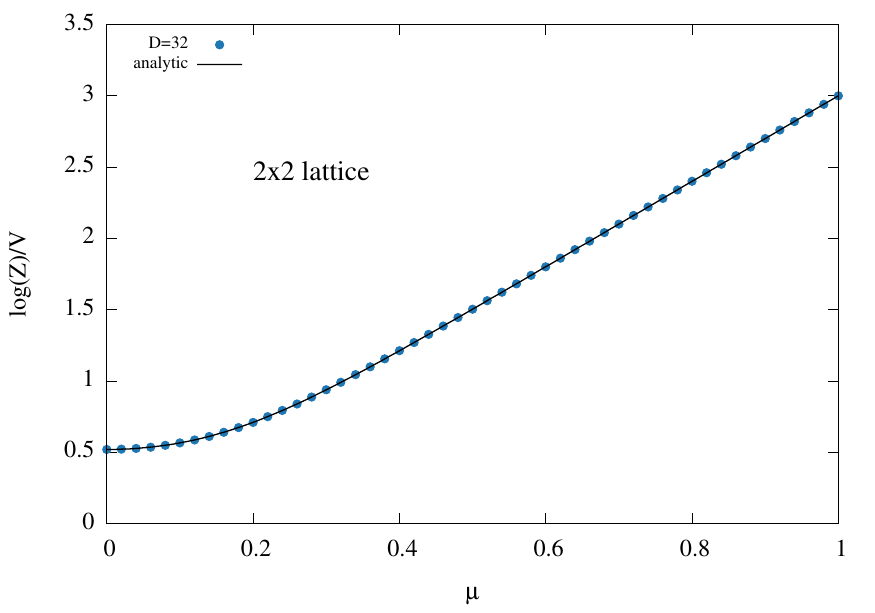}
\includegraphics[width=0.499\hsize]{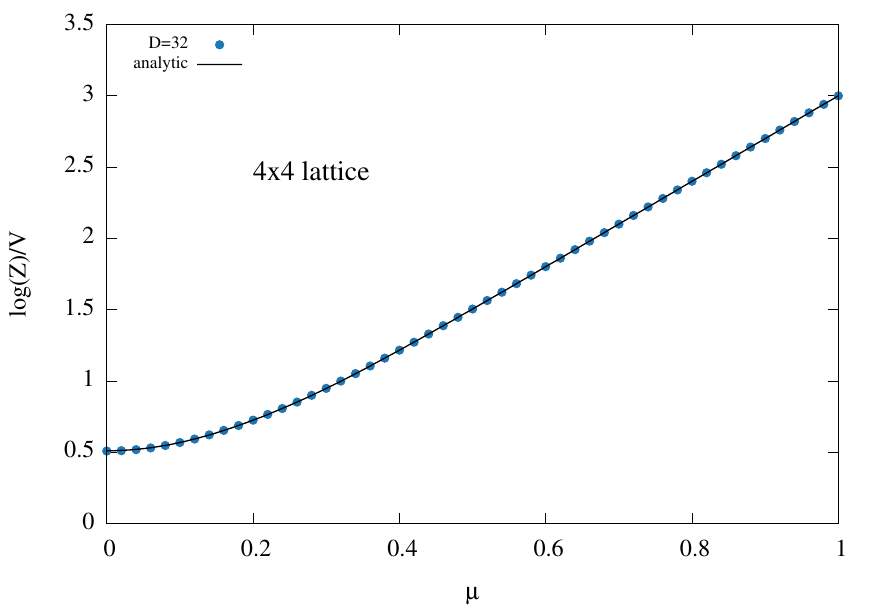}
}
\caption{Comparison of $\log(Z)/V$ versus $\mu$ for the baryon-only system on a $2\times2$ (left) and on a $4\times4$ (right) lattice computed using GHOTRG with D=32 and the analytic formulas \eqref{Z2x2bar} and \eqref{Z4x4bar}. }
\label{fig:baronly}
\end{figure}

We computed $\log(Z)/V$ as a function of the chemical potential $\mu$ on lattices of sizes $2\times2$, $2\times4$, $4\times2$, $4\times4$, $4\times8$ and $8\times4$ using GHOTRG with $D=32$ and compared these results with the analytical predictions given in \ref{app:analytic}. We find very good agreement between the numerical and analytical results. This is illustrated in Fig.~\ref{fig:baronly} for the $2\times2$ and $4\times4$ cases. 
These results confirm that the global minus signs that appear in the standard baryon-loop formulation of the model \cite{Karsch:1988zx} are correctly taken into account by the GHOTRG procedure.

\subsection{Meson-baryon partition function}

\begin{figure}
\centerline{
\includegraphics[width=0.499\hsize]{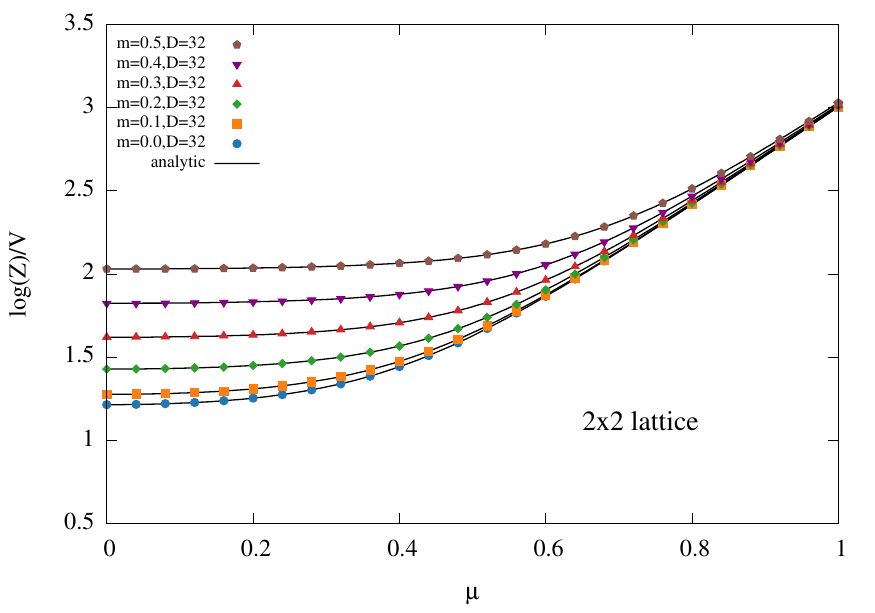}
\includegraphics[width=0.499\hsize]{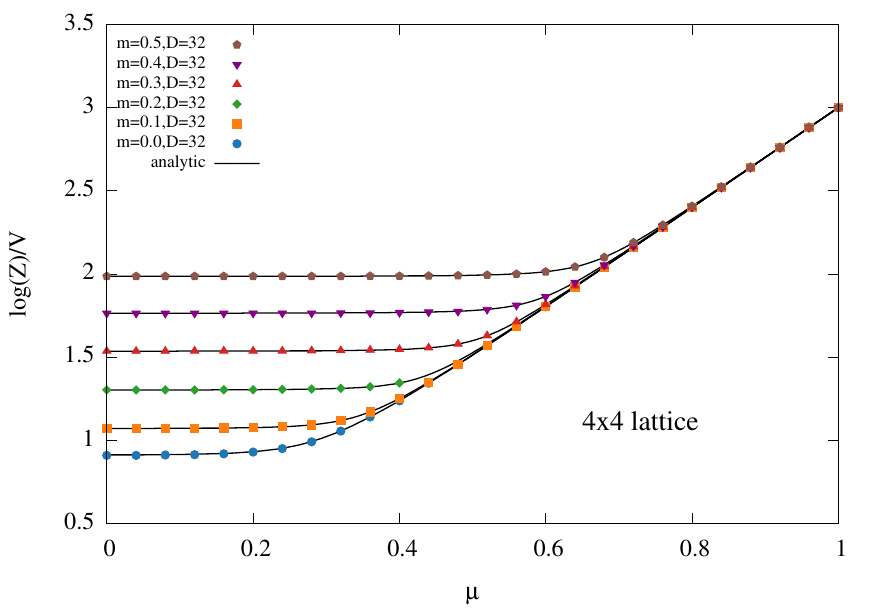}
}
\caption{Comparison of the full partition function $\log(Z)/V$ versus $\mu$ for a $2\times2$ lattice (left) and a $4\times4$ lattice (right) for masses $m \in \{0, 0.1, 0.2, 0.3, 0.4, 0.5\}$, computed using GHOTRG with D=32 and the analytic formulas \eqref{Z2x2} and \eqref{Z4x4}.}
\label{fig:full}
\end{figure}

Next we consider the full meson-baryon system of strong-coupling QCD in two dimensions, described by the tensor network \eqref{ZTNjf}. 
In Fig.~\ref{fig:full} we show $\log(Z)/V$ as a function of the chemical potential for a $2\times2$ and a $4\times4$ lattice with $m\in\{0,0.1,0.2,0.3,0.4,0.5\}$, computed using GHOTRG with fixed $D=32$, and compare with the analytic formulas of \ref{app:analytic}. We find very good agreement between the GHOTRG results and the exact values. 

\begin{figure}
\centerline{
\includegraphics[width=0.499\hsize]{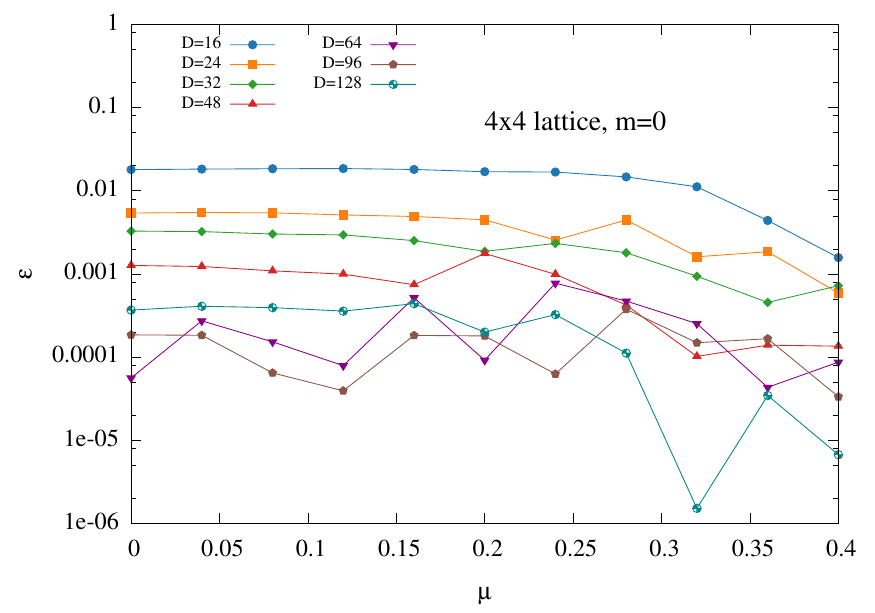}
\includegraphics[width=0.499\hsize]{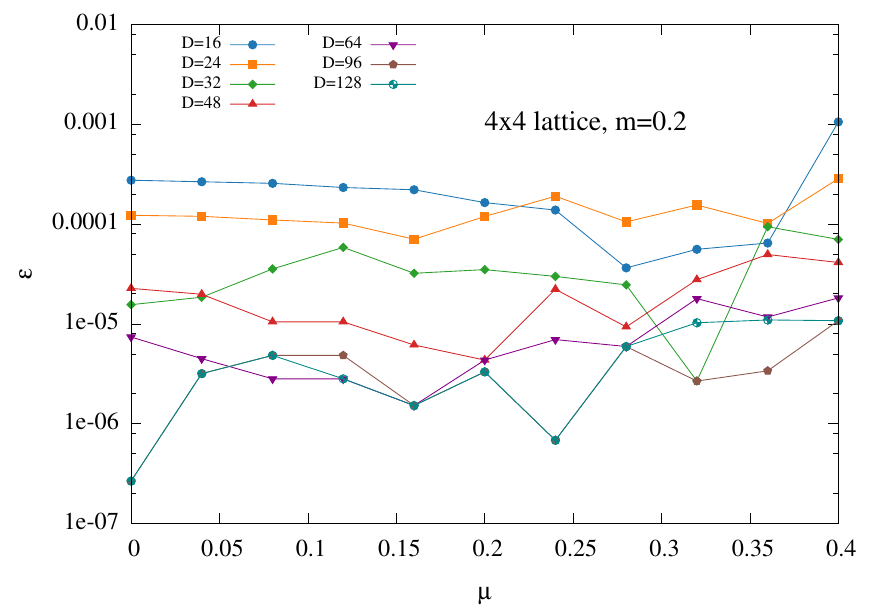}
}
\caption{Relative error $\epsilon$ on $\log(Z)/V$ versus $\mu$ for various values of $D$ on a $4\times4$ lattice for $m=0$ (left) and $m=0.2$ (right).}
\label{fig:conv}
\end{figure}

To verify the accuracy of the GHOTRG results, we show their relative deviation 
\begin{align}
\epsilon=\left|\frac{\log Z_\text{num}}{\log Z_\text{exact}} - 1\right| 
\label{epsilon}
\end{align}
from the exact result on a $4\times4$ lattice for $m=0$ and $m=0.2$ as a function of $\mu$ for various bond dimensions $D$ in Fig.\ \ref{fig:conv}. As expected, the accuracy typically improves with increasing $D$, but the behavior does not hold for all $\mu$ and $D$, which is related to the small size of the lattice, see below. For $m=0$ (left plot) the error is about a factor of 10 larger than for the nonzero mass $m=0.2$ (right plot). This shows that the tensor method is more accurate for larger masses, as is the case with most other simulation methods. Nevertheless, even in the chiral limit ($m=0$), the tensor method gives very satisfying results for this two-dimensional system.

\begin{figure}
\centerline{
\includegraphics[width=0.499\hsize]{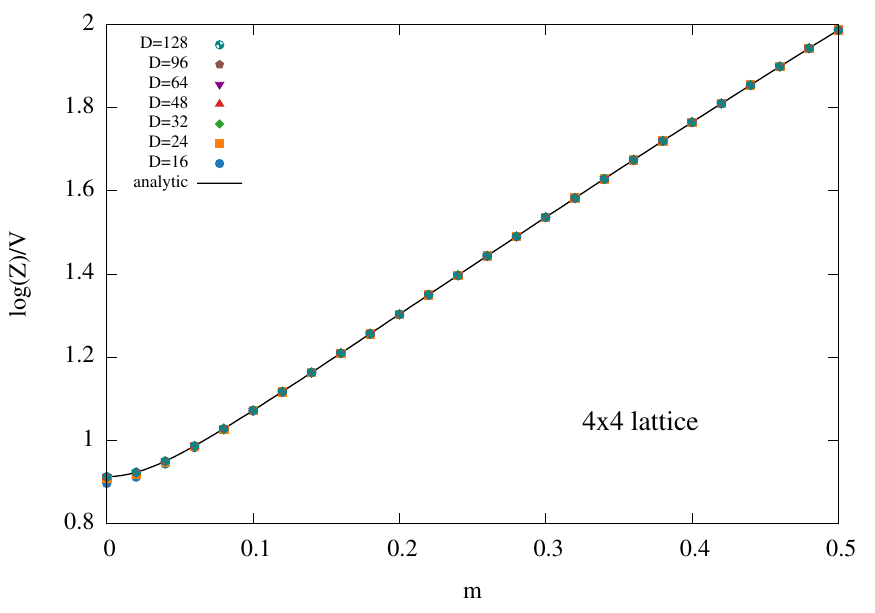}
\includegraphics[width=0.499\hsize]{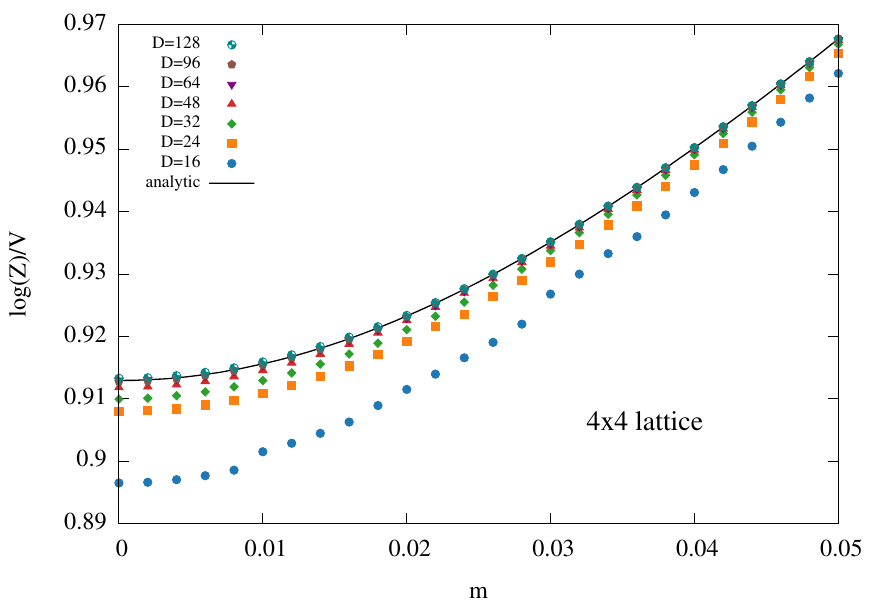}
}
\centerline{
\includegraphics[width=0.499\hsize]{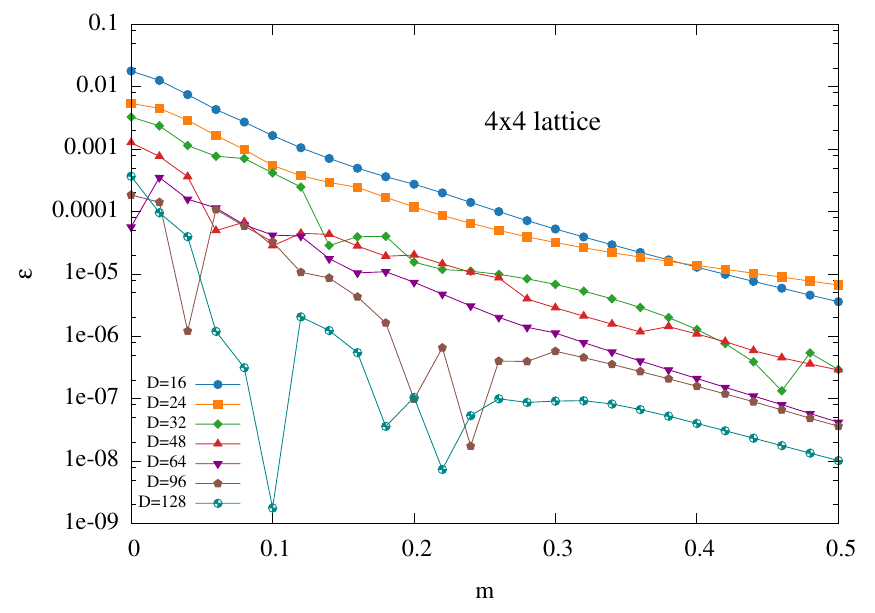}
\includegraphics[width=0.499\hsize]{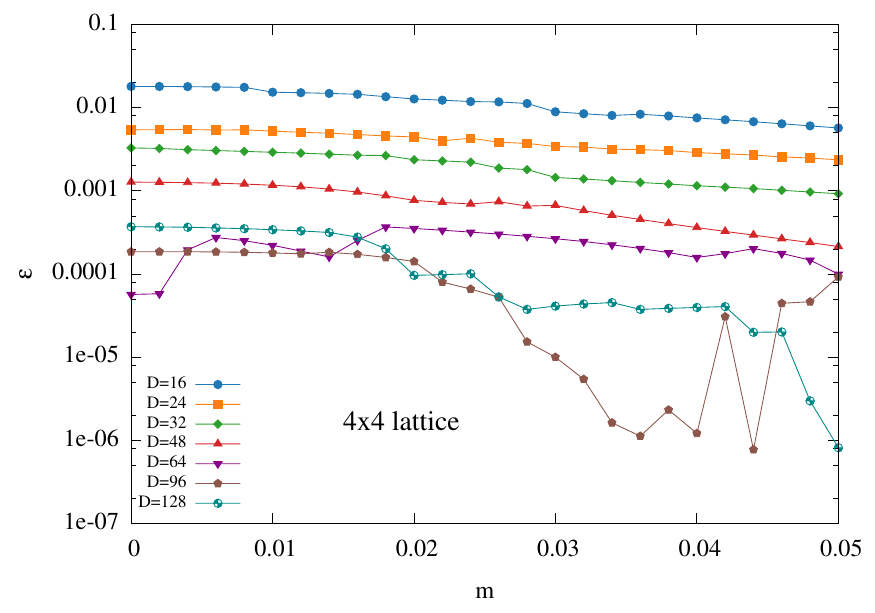}
}
\caption{Top row: $\log(Z)/V$ versus mass $m$ for $\mu=0$, computed using the GHOTRG method  with bond dimensions $D$ up to $D=128$ together with the exact values obtained from the analytic formula \eqref{Z4x4}. On the right we zoom in on the small-$m$ region, where the tensor method requires larger values of $D$.
Bottom row: relative error $\epsilon$, see \eqref{epsilon}, for the data shown in the top row. We see that the accuracy improves as $D$ becomes larger and the results converge to the exact values.}
\label{fig:Zvsm44}
\end{figure}

We also compute the mass dependence of $\log(Z)/V$ at zero chemical potential and verify our results with the analytic expression \eqref{Z4x4} on a $4\times 4$ lattice, see Fig.~\ref{fig:Zvsm44}. 
From the relative deviation $\epsilon$, shown in the bottom row of the figure, we observe that the accuracy improves as $D$ becomes larger and the results converge to the exact values. However, larger values of $D$ are required to get accurate results for smaller masses.

\begin{figure}
\centerline{
\includegraphics[width=0.499\hsize]{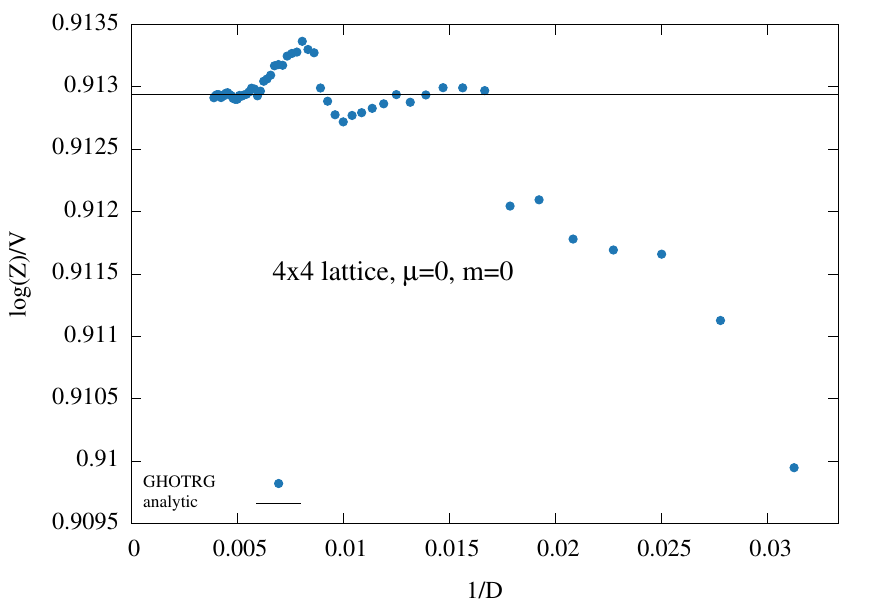}
\includegraphics[width=0.499\hsize]{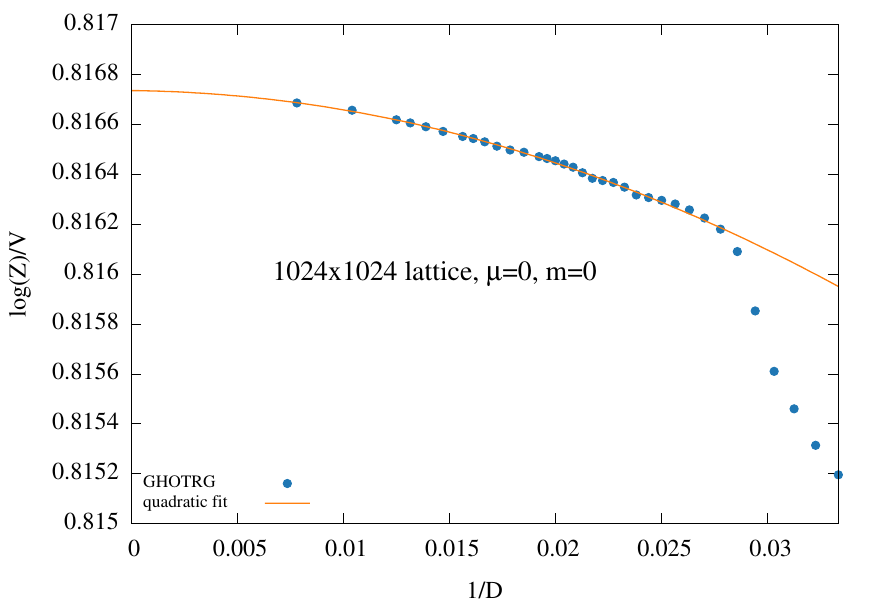}
}
\caption{Convergence of $\log(Z)/V$ versus bond dimension $D$ for $\mu=m=0$. Left: results for a $4\times4$ lattice (up to $D=256$) together with the exact value. Right: results for a $1024\times1024$ lattice (up to $D=128$) and quadratic fit in $1/D$ over the range $40\leq D\leq 128$.}
\label{fig:convZvsD}
\end{figure}

In Fig.~\ref{fig:convZvsD} we show a convergence study of $\log(Z)/V$ with respect to the bond dimension $D$ for $m=\mu=0$ on $4\times4$ and $1024\times1024$ lattices. The convergence behavior is quite erratic on the small lattice. Even though the accuracy is very good when $D$ is sufficiently large, the convergence is far from being monotonous.  For the large lattice, the convergence is much more stable, and a quadratic fit in $1/D$ allows us to make an extrapolation to $D\to\infty$.

\begin{figure}
\centerline{
\includegraphics[width=0.49\hsize]{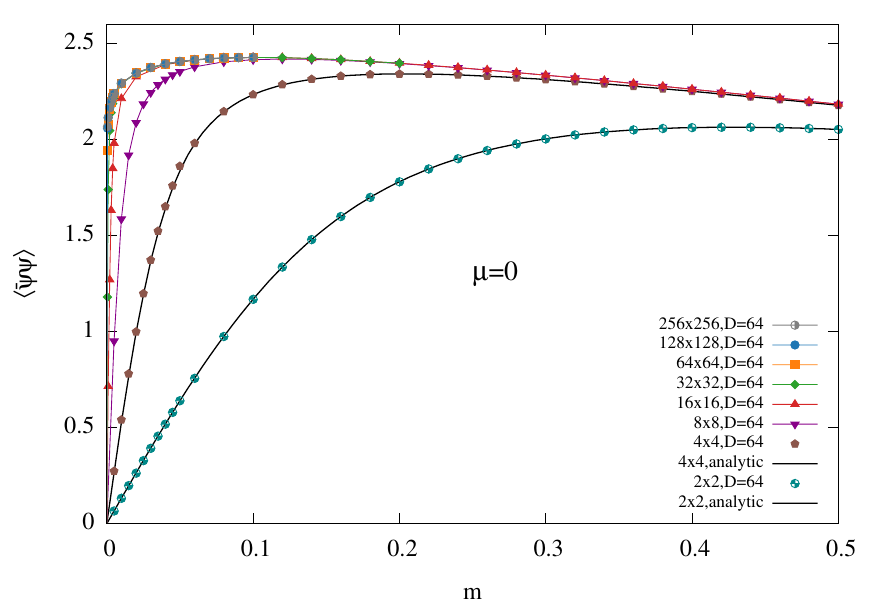}
\includegraphics[width=0.49\hsize]{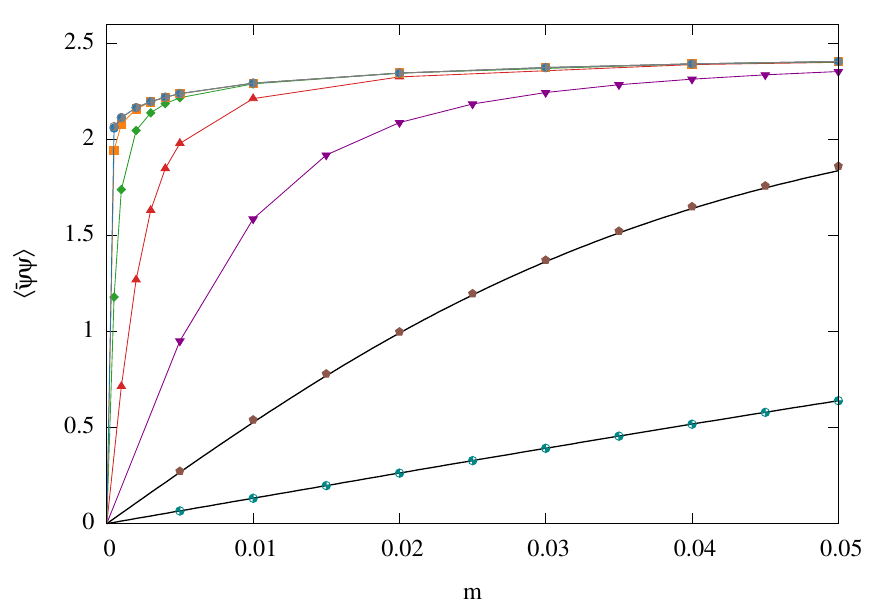}
}
\vspace{-3mm}
\caption{Chiral condensate versus mass for various volumes at $\mu=0$.  In the right plot we zoom in on the region close to $m=0$. As the mass gets smaller, larger volumes are required to reach the $V\to\infty$ limit.}
\label{fig:ccm}
\vspace{5mm}
\centerline{\includegraphics[width=0.47\hsize]{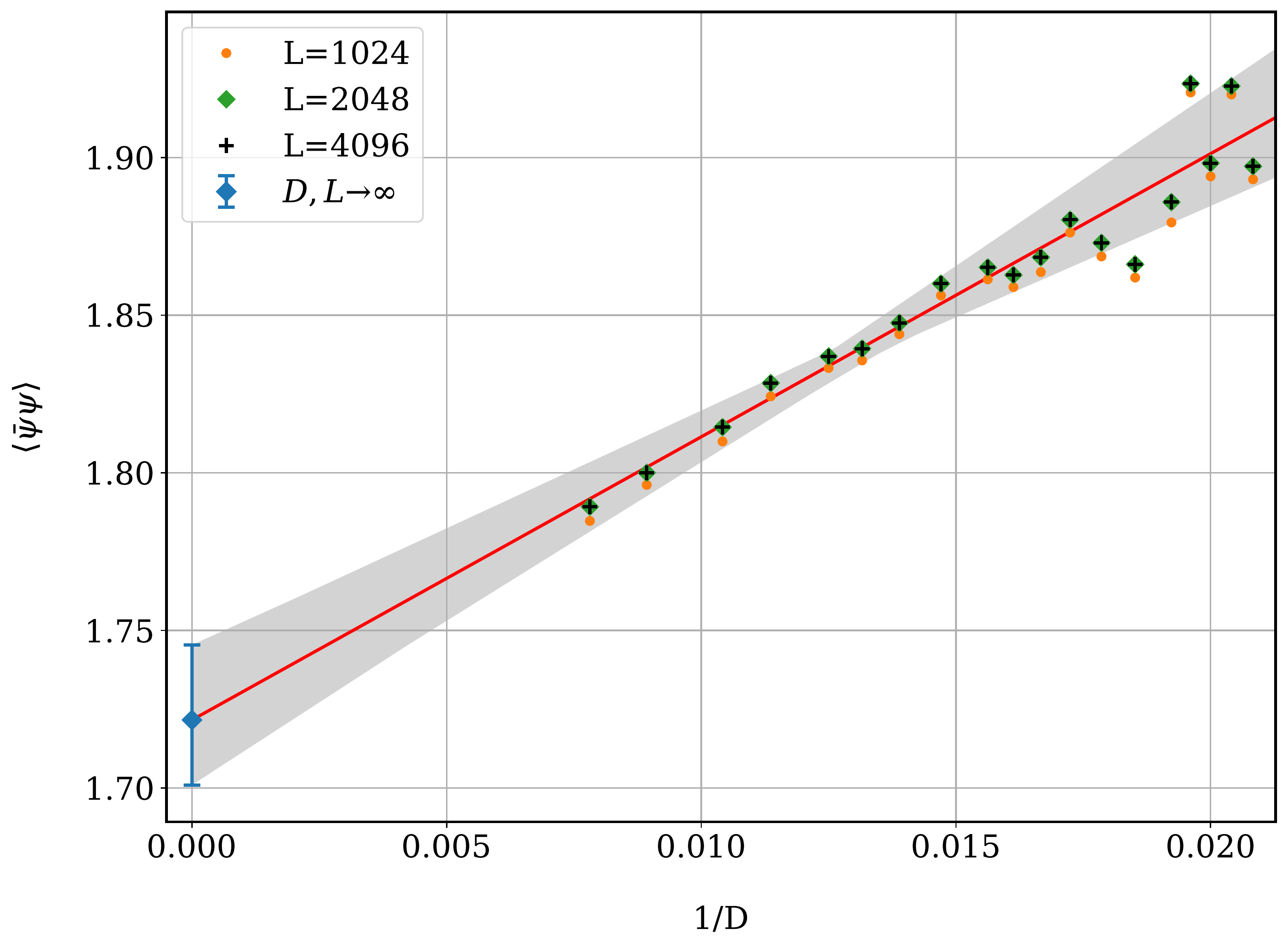}}
\vspace{-3mm}
\caption{Extrapolation of $\braket{\bar\psi\psi}$ to $D\to\infty$ for $m=10^{-5}$ and $L=4096$. The red line shows a linear fit in $1/D$ using all points with $D\geq48$, and the enclosing grey band includes $99\%$ of the linear fits using bootstrap sampling. We use the infinite-$D$ extrapolation for $L=4096$ as an estimate for $\lim_{V\to\infty}\lim_{D\to\infty}\braket{\bar\psi\psi}$, since the results do not change for $L\geq 2048$.
}
\label{bootstrap}
\vspace{5mm}
\centerline{
\includegraphics[width=0.49\hsize]{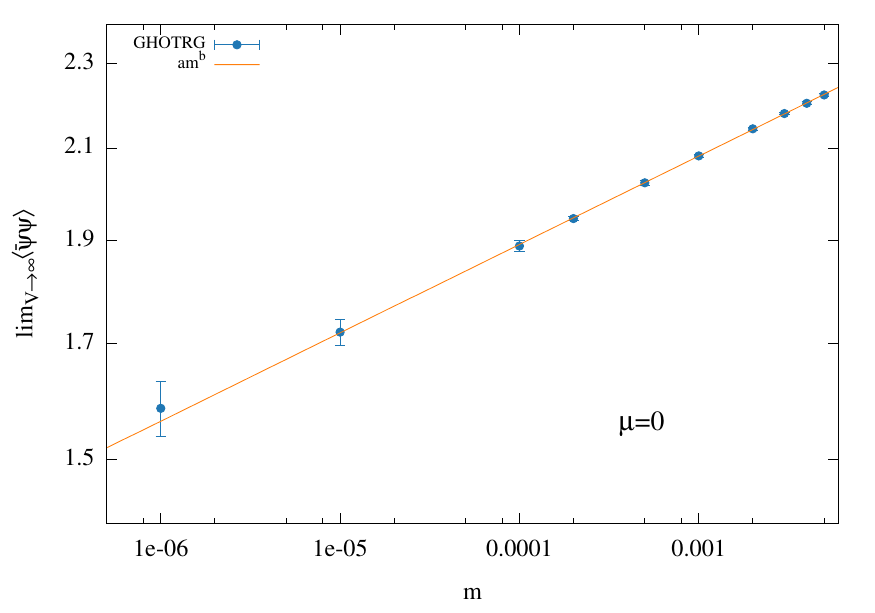}
\includegraphics[width=0.49\hsize]{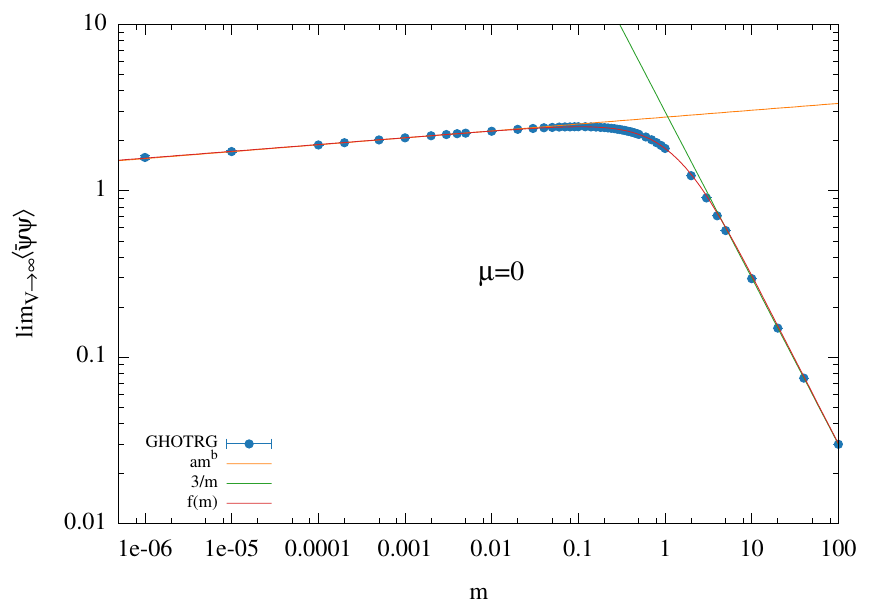}
}
\vspace{-3mm}
\caption{Infinite volume results for the chiral condensate as a function of the mass (for $\mu=0$). The error bars show estimates for the systematic errors of the extrapolations to infinite $D$ and infinite $V$. For small $m\leq 0.005$ the results lie on a straight line in a log-log plot and are thus well fitted by $\lim_{V\to\infty}\braket{\bar\psi\psi} = am^b$ (left panel). In the right panel we show a large mass interval where the results are fitted by $f(m)$, see text.}
\label{fig:ccmextrap}
\end{figure}

\subsection{Chiral condensate at zero chemical potential}

After validating the GHOTRG method for small lattices, where analytical results are available, we now consider larger lattices of size $L\times L$. First we compute the chiral condensate \eqref{cc} as a function of the mass and lattice volume at zero chemical potential. The aim is to investigate if the chiral symmetry is dynamically broken in this two-dimensional theory. To this end we look at the zero-mass and infinite-volume limit of the chiral condensate
\begin{align}
\lim_{m\to0}\lim_{V\to\infty}\braket{\bar\psi\psi} 
\begin{cases}
= 0 \qquad \text{no DCSB} ,
\\
\neq 0 \qquad \text{DCSB} ,
\end{cases}
\end{align}
where the order of the limits is crucial.

In Fig.\ \ref{fig:ccm} we show the evolution of the chiral condensate as a function of the mass for various lattice sizes at fixed $D=64$. Although these results are computed at fixed $D$, they already illustrate how the chiral condensate converges to its infinite volume limit for the different mass values. As the mass gets smaller, larger volumes are needed to approach this limit. 

We now perform a detailed analysis of the chiral condensate by extrapolating to $D\to\infty$ for each mass and volume, and then extrapolating this infinite-$D$ result to $V\to\infty$ for each mass value. 
An example for such an extrapolation, together with its error estimate, is shown in Fig.\ \ref{bootstrap} for $m=10^{-5}$. We observe that the lattice size needed to obtain an estimate for the $V\to\infty$ limit increases with decreasing mass.
The results of these extrapolations as a function of the mass are shown in Fig.~\ref{fig:ccmextrap}. For small $m<0.005$, the results lie on a straight line in a log-log plot and are thus well fitted by $\lim_{V\to\infty}\braket{\bar\psi\psi} = am^b$. For the fit shown in Fig.~\ref{fig:ccmextrap} (left), the fit parameters are given by $a=2.77$, $b=0.0414$. This shows that the chiral symmetry is not dynamically broken in two-dimensional strong-coupling QCD with (two tastes of) staggered quarks. 

For large masses, the chiral condensate is asymptotically given by $3/m$ at leading order, which can easily be derived from the partition function \eqref{Z}. We therefore fit the infinite-volume limit of the chiral condensate over the full mass range by the empirical formula $f(m)=(am^b+cm)/(1+dm+(c/3)m^2)$, which interpolates between the asymptotic behaviors, see Fig.~\ref{fig:ccmextrap} (right). The fitted parameter values are $a=2.77$, $b=0.0409$, $c=1.05$, $d=0.770$.

\subsection{Particle number density and chiral condensate at nonzero chemical potential}

\begin{figure}
\centerline{
\includegraphics[width=0.499\hsize]{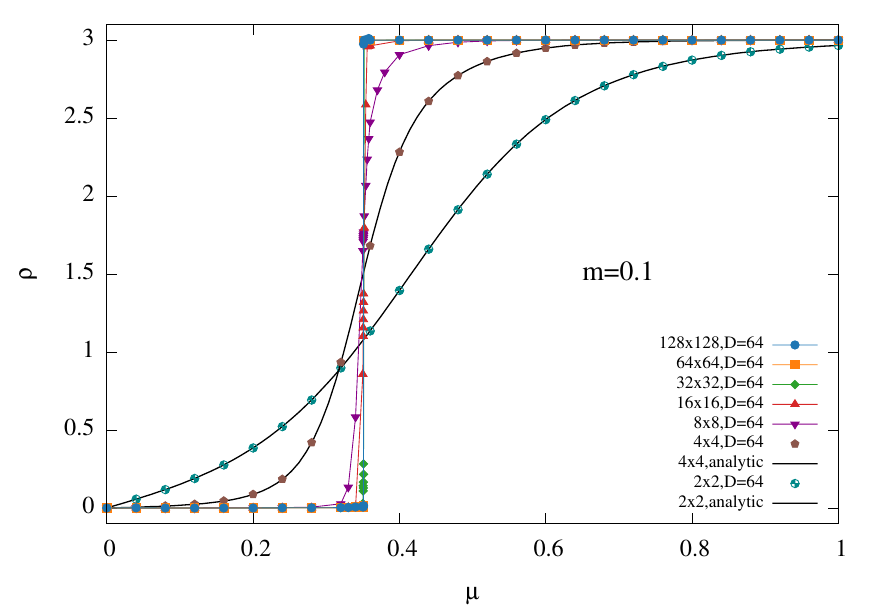}
\includegraphics[width=0.499\hsize]{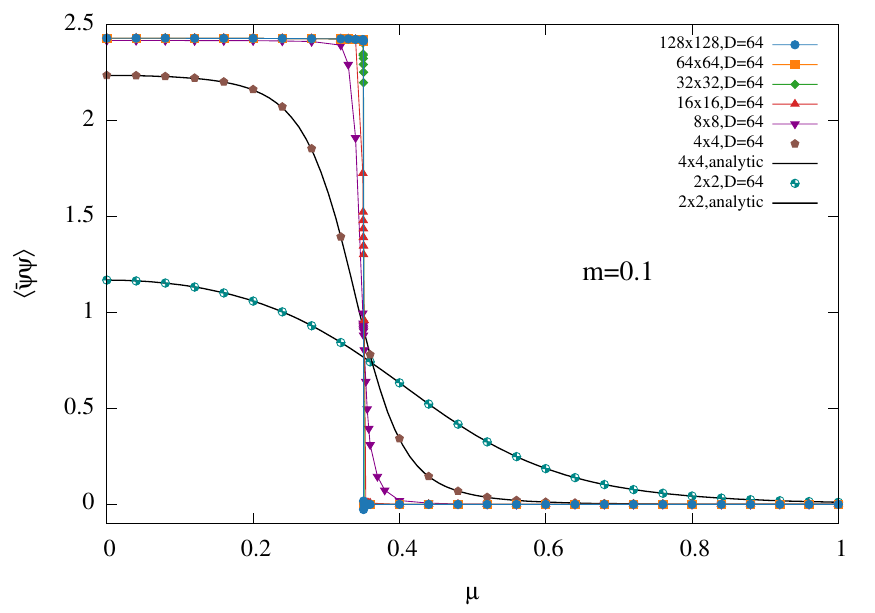}
}
\caption{Quark number density $\rho$  (left) and chiral condensate $\braket{\bar\psi\psi}$ (right) versus $\mu$ for $m=0.1$. Both were computed for various lattice sizes with $D=64$. We observe some evidence for a first-order phase transition at $\mu_c\approx0.3508$.}
\label{fig:n_m=0.1}
\end{figure}

Finally, we use the GHOTRG method to investigate the behavior of the model at nonzero chemical potential. 
For $m=0.1$ we study the quark number density \eqref{rho} and the chiral condensate \eqref{cc} as a function of the chemical potential, for lattice sizes up to $L=128$. The GHOTRG results for $D=64$ are shown in Fig.~\ref{fig:n_m=0.1}. For small lattices ($2\times2$ and $4\times4$), they agree well with the exact values obtained from the analytic formulas of \ref{app:analytic}. For larger lattices, the results quickly converge to the $V\to\infty$ limit. There appears to be a first-order phase transition around $\mu_c\approx0.3508$.  Above this critical chemical potential, we observe both a nonzero quark number density $\rho$ and a restoration of the chiral symmetry, i.e., $\braket{\bar\psi\psi}=0$.

\section{Conclusions}
\label{sec:conclusions}

In this paper we developed a tensor-network renormalization group framework, based on the GHOTRG method,  specifically tailored for strongly-coupled two-dimensional QCD with staggered quarks. In its dual formulation, the partition function is composed of mesonic and baryonic degrees of freedom. The Grassmann variables in the baryonic contributions to the partition function cannot be integrated out without introducing non-local sign factors. Therefore, the partition function cannot be written as a full contraction of a numeric tensor network and the standard HOTRG method cannot be applied. However, this problem can be resolved by constructing a tensor network consisting of local numeric and Grassmann tensors. When the lattice is then coarsened during the renormalization group procedure, the blocking of two adjacent sites now consists of two steps: First the Grassmann tensors on the two adjacent sites are blocked, yielding a new Grassmann tensor on the coarse lattice. This procedure generates a local sign factor,
which is absorbed in the contraction of the numeric tensors on the two adjacent sites. Just as in standard HOTRG, the coarse numeric tensor is then subjected to an HOSVD approximation to avoid an exponential increase of its dimensionality. After each renormalization group step, the partition function is represented by a coarsened tensor network of local tensors that are again products of a numeric and a Grassmann tensor. At each blocking step, the number of Grassmann variables is reduced by a factor of two and the HOSVD procedure reduces the dimensions of the fat indices of the numeric tensor back from $D^2$ to $D$.
This procedure is repeated until the whole lattice has been reduced to a single site and the partition function can be computed, taking into account the boundary conditions.

Our version of the GHOTRG procedure allows for a tensor-network computation of the  partition function with a computational cost that is similar to that of standard HOTRG. 
This can be achieved since the Gram matrices, used in the construction of the truncation matrices, are block diagonal in the Grassmann parity. Without this block-diagonal structure, truncations from dimension $4D^2$ to $D$ would be required at each blocking step to keep the dimensions of the coarse local tensors under control, as is the case in applications of the GHOTRG method for some other fermionic problems \cite{Sakai:2017jwp}. This would make the GHOTRG method substantially more expensive.

We have validated our version of the GHOTRG method by comparing with exact results on small lattices. On large lattices, we have studied the chiral condensate as a function of the mass and the volume at zero chemical potential, and showed that the chiral symmetry is not dynamically broken in this two-dimensional model in the chiral limit ($m\to0$). At nonzero chemical potential we computed both the quark number density and the chiral condensate and found some evidence for a first-order phase transition to a phase with nonzero density where the chiral symmetry is restored (for nonzero mass).

In future work we will apply the method to strong-coupling QCD in higher dimensions and also extend it beyond the infinite-coupling limit.

\section*{Acknowledgements}

We thank Aaron Holmer, Christina Jäger, Marco Lents, Pascal Milde and Thomas Samberger for useful discussions.

\appendix 

\section{Commutativity of $\cK^{(\X)}$}
\label{app:commuting}

Let us assume that the entries $T_{\j_{x,-1}\j_{x,1}\j_{x,-2}\j_{x,2}}$ of the numeric tensor vanish unless the Grassmann parities $f\equiv f(j)$ obey
\begin{align}
(f_{x,-2}+f_{x,-1}+f_{x,2}+f_{x,1}) \!\!\!\mod 2 &= 0 ,
\label{fmod2}
\end{align}
which is satisfied by the initial local tensor \eqref{initT} due to $\delta_{x\in{\cal M}}$ or $\delta_{x\in{\cal B}}$, see \eqref{xinM} and \eqref{xinB}.
The contraction \eqref{defcM} of adjacent tensors in the \dir1-direction, obtained by a sum over $j_{x,1}$,  yields a tensor which only has nonzero entries for
\begin{align}
(\f_{x,-2}+\f_{x,-1}+\f_{x,2} +
\f_{x+\hat1,-2}+\f_{x+\hat1,2}+\f_{x+\hat1,1}) \!\!\!\mod 2 &= 0 .
\label{condition2}
\end{align}
For $f_{x,1}=0$ the sum of the first three and last three terms are both even, while for $f_{x,1}=1$ they are both odd. Therefore the Grassmann tensor $\cG^{(\X)}$ of \eqref{Kx1a} can be considered to be Grassmann-even in the partition function \eqref{ZMG}.

Using \eqref{fatmode}, the condition \eqref{condition2} can be replaced by 
\begin{align}
(\cf_{X,-2}+\cf_{X,2}+\f_{x,-1}+\f_{x+\hat1,1}) \!\!\!\mod 2 = 0 .
\label{comid}
\end{align}
This means the tensor entries of $\cS$ of \eqref{defcT} are zero whenever \eqref{comid} is not satisfied.
On the other hand, when this condition is satisfied, the coarse-lattice Grassmann tensors $\cK^{(\X)}$ of \eqref{Kbarx3} are commuting in the partition function since the same argument applies for all $\X$.

When applying HOSVD, see Sec.~\ref{sec:applyU}, the Grassmann parity $\cg$ of the new indices $\cj$ corresponds to $\cf$ because of the block-diagonal nature of the truncation matrices, such that the condition is replaced by 
\begin{align}
(\cg_{X,-2}+\cg_{X,2}+\f_{x,-1}+\f_{x+\hat1,1}) \!\!\!\mod 2 = 0 .
\label{comidg}
\end{align}

Equation \eqref{comidg} is the equivalent of \eqref{fmod2} on the coarse lattice, see \eqref{renaming}. A completely analogous argument applies for contractions in the \dir2-direction, see Sec.~\ref{sec:contract2}.
Therefore the Grassmann tensors can be considered to be commuting throughout the whole blocking procedure.

\section{Implementation}
\label{app:implementation}

Below we discuss implementation details of our GHOTRG method.
In standard HOTRG, the HOSVD approximation of the coarse-lattice tensor is typically performed without explicitly constructing the latter, for reasons of computational and storage efficiency. In GHOTRG, the blocking of the Grassmann tensors introduces additional sign factors in the coarse-lattice numerical tensor, see \eqref{defcT}, and hence, we modify the standard HOTRG procedure accordingly.

\subsection{Computing the semi-orthogonal truncation matrices $U$}
\label{app:Q}

The HOSVD procedure requires the computation of the left singular vectors for the matrizations $M$ of the coarse-lattice tensor $\cS$ of \eqref{defcT}, with respect to its fat links.
To avoid the explicit construction of the coarse-lattice tensor $\cS$, we compute these singular vectors as eigenvectors of the corresponding Gram matrices $Q=M M^T$.
We adapt the calculation of $Q$ in the standard HOTRG method to include the additional sign factors coming from the Grassmann integrations. 

For a contraction in the \dir1-direction, the matrization with respect to the backward \dir2-direction leads to the Gram matrix
\begin{align}
\lefteqn{Q^{-}_{(\j_{x,-2},\j_{x+\hat1,-2})\,,\,(\jp_{x,-2},\jp_{x+\hat1,-2})}}
\notag\\
&\hspace{10ex}= \sum_{\j_{x,-1},\j_{x+\hat1,1},\j_{x,2},\j_{x+\hat1,2}} 
\cS^{(\X)}_{\j_{x,-1}\j_{x+\hat1,1}(\j_{x,-2},\j_{x+\hat1,-2})(\j_{x,2},\j_{x+\hat1,2})} 
\cS^{(\X)}_{\j_{x,-1}\j_{x+\hat1,1}(\jp_{x,-2},\jp_{x+\hat1,-2})(\j_{x,2},\j_{x+\hat1,2})} \notag\\
&\hspace{10ex}= \sum_{\j_{x,-1},\j_{x+\hat1,1},\j_{x,2},\j_{x+\hat1,2}}
\Bigg(
\sigma_{\f_{x,2}\f_{x+\hat1,-2}\f_{x+\hat1,2}} 
\sum_{\j_{x,1}} 
T^{(x)}_{\j_{x,-1}\j_{x,1}\j_{x,-2}\j_{x,2}}
T^{(x+\hat1)}_{\j_{x,1}\j_{x+\hat1,1}\j_{x+\hat1,-2}\j_{x+\hat1,2}}
\Bigg) \notag\\
&\hspace{30ex} \times \Bigg(
\sigma_{\f_{x,2}\f'_{x+\hat1,-2}\f_{x+\hat1,2}} 
\sum_{\jp_{x,1}} 
T^{(x)}_{\j_{x,-1}\jp_{x,1}\jp_{x,-2}\j_{x,2}}
T^{(x+\hat1)}_{\jp_{x,1}\j_{x+\hat1,1}\jp_{x+\hat1,-2}\j_{x+\hat1,2}}
\Bigg) .
\label{Qmin}
\end{align}
Note that the Grassmann parities $\f$ in the sign factors $\sigma$ are functions of the corresponding indices $j$.

To improve the efficiency of the computation of $Q^-$ and to reduce the required storage,
we would like to reshuffle the factors in the previous expression, such that the tensors $T$ at the same positions are contracted first, as is usually done in standard HOTRG. The additional couplings between the tensors, caused by the sign factors, complicate the reordering of the product. 

As the nonzero entries of $M^-$ satisfy \eqref{parcons}, the nonzero entries of $Q^-$ have Grassmann parities satisfying
\begin{align}
(f_{x,-2}+f_{x+\hat1,-2})\!\!\!\!\mod2 = (f'_{x,-2}+f'_{x+\hat1,-2})\!\!\!\!\mod2 ,
\end{align}
such that $Q^-$ is block diagonal with the nonzero blocks being either even-even or odd-odd blocks in $f_{x,-2}+f_{x+\hat1,-2}$ and $f'_{x,-2}+f'_{x+\hat1,-2}$. 

The product of the sign factors $\sigma$ in \eqref{Qmin} simplifies to 
\begin{align}
\sigma_{\f_{x,2}\f_{x+\hat1,-2}\f_{x+\hat1,2}} 
\sigma_{\f_{x,2}\f'_{x+\hat1,-2}\f_{x+\hat1,2} 
}
&= (-1)^{\f_{x,2}(\f_{x+\hat1,-2}+\f_{x+\hat1,2})}
(-1)^{\f_{x,2}(\f'_{x+\hat1,-2}+\f_{x+\hat1,2})}
\notag\\
&= (-1)^{\f_{x,2}(\f_{x+\hat1,-2}+\f'_{x+\hat1,-2})}
= \sigma_{\f_{x,2}\f_{x+\hat1,-2}\f'_{x+\hat1,-2}} .
\end{align}
We now reorder the sums in \eqref{Qmin} such that the tensors on equal sites can be contracted first.

The sums over the indices that only appear in the two factors of $T^{(x)}$ yields
\begin{align}
A^-_{\j_{x,-2}\jp_{x,-2}\j_{x,1}\jp_{x,1}\f}
&= \sum_{\j_{x,-1},\j_{x,2}^{(\f)}} 
T^{(x)}_{\j_{x,-1}\j_{x,1}\j_{x,-2}\j_{x,2}^{(\f)}}
T^{(x)}_{\j_{x,-1}\jp_{x,1}\jp_{x,-2}\j_{x,2}^{(\f)}}
,\qquad \text{for $f=0,1$,}
\end{align}
where the indices $j^{(0)}$ and $j^{(1)}$ in the sum denote the indices $j$ with even and odd Grassmann parities, respectively.
The storage and computational costs scale as $\Mem \propto 2D^4$ and $\Comp \propto D^6$, respectively. 
Note that we cannot add the $\f_{x,2}=0$ and $\f_{x,2}=1$ contributions when constructing the auxiliary tensor $A^-$ since $\f_{x,2}$ also appears in $\sigma_{\f_{x,2}\f_{x+\hat1,-2}\f'_{x+\hat1,-2}}$.
Analogously, for the two factors $T^{(x+\hat1)}$, we construct 
\begin{align}
B^-_{\j_{x+\hat1,-2}\jp_{x+\hat1,-2}\j_{x,1}\jp_{x,1}}
&= \sum_{\j_{x+\hat1,1},\j_{x+\hat1,2}}
 T^{(x+\hat1)}_{\j_{x,1}\j_{x+\hat1,1}\j_{x+\hat1,-2}\j_{x+\hat1,2}} 
 T^{(x+\hat1)}_{\jp_{x,1}\j_{x+\hat1,1}\jp_{x+\hat1,-2}\j_{x+\hat1,2}}
\label{Bb}
\end{align}
with $\Mem \propto D^4$ and $\Comp \propto D^6$. Finally we compute
\begin{align}
Q^{-}_{(\j_{x,-2},\j_{x+\hat1,-2})\,,\,(\jp_{x,-2},\jp_{x+\hat1,-2})}
= \sum_{\f_{x,2}} \sigma_{\f_{x,2}\f_{x+\hat1,-2}\f'_{x+\hat1,-2}}
\sum_{\j_{x,1},\jp_{x,1}} A^-_{\j_{x,-2}\jp_{x,-2}\j_{x,1}\jp_{x,1}\f_{x,2}} B^-_{\j_{x+\hat1,-2}\jp_{x+\hat1,-2}\j_{x,1}\jp_{x,1}}
\end{align}
with $\Mem \propto 2D^4$ and $\Comp \propto 2D^6$, by first contracting $A^-$ and $B^-$ and then summing over $f_{x,2}$.

Let us now look at the forward \dir2-direction, which is slightly different because of the backward-forward asymmetry of the sign factor:
\begin{align}
\lefteqn{Q^{+}_{(\j_{x,2},\j_{x+\hat1,2})\,,\,(\jp_{x,2}\jp_{x+\hat1,2})}}
\notag\\
&\hspace{7ex}= \sum_{\j_{x,-1},\j_{x+\hat1,1},\j_{x,-2},\j_{x+\hat1,-2}}
\Bigg(\sigma_{\f_{x,2}\f_{x+\hat1,-2}\f_{x+\hat1,2}}
\sum_{\j_{x,1}} T^{(x)}_{\j_{x,-1}\j_{x,1}\j_{x,-2}\j_{x,2}} T^{(x+\hat1)}_{\j_{x,1}\j_{x+\hat1,1}\j_{x+\hat1,-2}\j_{x+\hat1,2}} 
\Bigg)\notag\\
&\hspace{29ex}\times \Bigg(\sigma_{\f'_{x,2}\f_{x+\hat1,-2}\f'_{x+\hat1,2}}
\sum_{\jp_{x,1}} T^{(x)}_{\j_{x,-1}\jp_{x,1}(lf)_{x,-2}\jp_{x,2}} T^{(x+\hat1)}_{\jp_{x,1}\j_{x+\hat1,1}(lf)_{x+\hat1,-2}\jp_{x+\hat1,2}} 
\Bigg) .
\end{align}
The product of sign factors yields
\begin{align}
\lefteqn{\sigma_{\f_{x,2}\f_{x+\hat1,-2}\f_{x+\hat1,2}}
\sigma_{\f'_{x,2}\f_{x+\hat1,-2}\f'_{x+\hat1,2}} = (-1)^{\f_{x,2}(\f_{x+\hat1,-2}+\f_{x+\hat1,2})}
(-1)^{\f'_{x,2}(\f_{x+\hat1,-2}+\f'_{x+\hat1,2})}}
\notag\\
 &\hspace{4ex}= 
(-1)^{\f_{x,2}\f_{x+\hat1,2}+\f'_{x,2}\f'_{x+\hat1,2}} (-1)^{\f_{x+\hat1,-2}(\f_{x,2}+\f'_{x,2})} 
= (-1)^{\f_{x,2}\f_{x+\hat1,2}+\f'_{x,2}\f'_{x+\hat1,2}} \sigma_{\f_{x+\hat1,-2}\f_{x,2}\f'_{x,2}} .
\end{align}

We introduce $A^+$ and $B^+$ for the sites $x$ and $x+\hat1$, respectively,
\begin{align}
A^+_{\j_{x,2}\jp_{x,2}\j_{x,1}\jp_{x,1}}
&= \sum_{\j_{x,-1},\j_{x,-2}} 
T^{(x)}_{\j_{x,-1}\j_{x,1}\j_{x,-2}\j_{x,2}} T^{(x)}_{\j_{x,-1}\jp_{x,1}\j_{x,-2}\jp_{x,2}} ,
\\
B^+_{\j_{x+\hat1,2}\jp_{x+\hat1,2}\j_{x,1}\jp_{x,1}\f} 
&= \sum_{\j_{x+\hat1,1},j_{x+\hat1,-2}^{(\f)}}  
T^{(x+\hat1)}_{\j_{x,1}\j_{x+\hat1,1}\j_{x+\hat1,-2}^{(\f)}\j_{x+\hat1,2}} T^{(x+\hat1)}_{\jp_{x,1}\j_{x+\hat1,1}\j_{x+\hat1,-2}^{(\f)}\jp_{x+\hat1,2}} ,\qquad \text{for $f=0,1$}
.
\label{Bplus}
\end{align}
The construction of $A^+$ has cost $\Mem \propto D^4$, $\Comp \propto D^6$ and $B^+$ has $\Mem \propto 2D^4$, $\Comp \propto D^6$.
The $Q^+$ matrix is then
\begin{align}
\lefteqn{Q^{+}_{(\j_{x,2},\j_{x+\hat1,2})\,,\,(\jp_{x,2},\jp_{x+\hat1,2})}}
\notag\\
&\hspace{2ex}= 
(-1)^{\f_{x,2}\f_{x+\hat1,2}+\f'_{x,2}\f'_{x+\hat1,2}} 
\sum_{\f_{x+\hat1,-2}}
\sigma_{\f_{x+\hat1,-2}\f_{x,2}\f'_{x,2}}
\sum_{\j_{x,1},\jp_{x,1}}
A_{\j_{x,2}\jp_{x,2}\j_{x,1}\jp_{x,1}}
B_{\j_{x+\hat1,2}\jp_{x+\hat1,2}\j_{x,1}\jp_{x,1}\f_{x+\hat1,-2}}
\label{Qf}
\end{align}
with memory cost $\Mem \propto 2D^4$ and computational cost $\Comp \propto 2D^6$.

\subsection{Truncating the coarse tensor}
\label{app:T}

We present an efficient implementation of the truncation of the coarse tensor $\cS$ for a contraction in the \dir1-direction, where the dimensions of the vector spaces corresponding to the \dir2-direction are reduced from $D^2$ to $D$ using the semi-orthogonal truncation matrix $U$ constructed in Sec.~\ref{sec:hosvd},
\begin{align}
\lefteqn{\cT_{\j_{x,-1}\j_{x+\hat1,1}\clf_{\X,-2}\clf_{\X,2}}
= \sum_{\j_{x,-2},\j_{x+\hat1,-2},\j_{x,2},\j_{x+\hat1,2}} U_{(\j_{x,-2},\j_{x+\hat1,-2})\cj_{\X,-2}} U_{(\j_{x,2},\j_{x+\hat1,2})\cj_{\X,2}}}
\notag\\
&\hspace{30ex}\times
\sigma_{\f_{x,2}\f_{x+\hat1,-2}\f_{x+\hat1,2}} 
\sum_{\j_{x,1}} 
T^{(x)}_{\j_{x,-1}\j_{x,1}\j_{x,-2}\j_{x,2}}
T^{(x+\hat1)}_{\j_{x,1}\j_{x+\hat1,1}\j_{x+\hat1,-2}\j_{x+\hat1,2}} 
\label{newT2d}
\end{align}
with 
\begin{align}
\sigma_{\f_{x,2}\f_{x+\hat1,-2}\f_{x+\hat1,2}} 
&= (-1)^{\f_{x,2}(\f_{x+\hat1,-2}+\f_{x+\hat1,2})} 
=(-1)^{\f_{x,2}\f_{x+\hat1,-2}} (-1)^{\f_{x,2}\f_{x+\hat1,2}} .
\end{align}
We again want to avoid the explicit construction of the coarse tensor $\cS$ and therefore reorganize the contractions.
For later use we define two tensors by making the following products of $U$ with the factorized sign factors,
\begin{align}
A^-_{\j_{x,-2}\j_{x+\hat1,-2}\clf_{\cx,-2}f_{x,2}} &= (-1)^{f_{x,2}f_{x+\hat1,-2}} U_{(\j_{x,-2},\j_{x+\hat1,-2})\clf_{\cx,-2}} ,
\\
A^+_{\j_{x,2}\j_{x+\hat1,2}\clf_{\cx,2}} &= (-1)^{f_{x,2}f_{x+\hat1,2}} U_{(\j_{x,2},\j_{x+\hat1,2})\clf_{\cx,2}}
\end{align}
with $\Mem \propto \Comp \propto 2D^3$ and $D^3$, respectively.

In the following, indices before the ``|" represent the indices of the outmost loops of the implementation. These indices are not explicitly present in the auxiliary tensors, which reduces the storage requirements of the computation.
We construct the further auxiliary tensors (note that $f_{x,2}\equiv f_{x,2}(j_{x,2})$ is not summed over),
\begin{align}
B_{\j_{x,-1}\clf_{\cx,-2}\,|\,\j_{x,1}\j_{x,2}\j_{x+\hat1,-2}} 
= \sum_{\j_{x,-2}} A^-_{\j_{x,-2}\j_{x+\hat1,-2}\clf_{\cx,-2}f_{x,2}}  T^{(x)}_{\j_{x,-1}\j_{x,1}\j_{x,-2}\j_{x,2}} 
\end{align}
with $\Mem \propto D^3$ and $\Comp \propto D^6$.
Then, we contract $B$ with the second $T$,
\begin{align}
C_{\j_{x,-1}\clf_{\cx,-2}\,|\,\j_{x+\hat1,1}\j_{x,2}\j_{x+\hat1,2}}
&= \sum_{\j_{x,1},\j_{x+\hat1,-2}} B_{\j_{x,-1}\clf_{\cx,-2}\,|\,\j_{x,1}\j_{x,2}\j_{x+\hat1,-2}} 
T^{(x+1)}_{\j_{x,1}\j_{x+\hat1,1}\j_{x+\hat1,-2}\j_{x+\hat1,2}}
\end{align}
with $\Mem \propto D^3$ and $\Comp \propto D^{7}$. 
Finally, we make the remaining tensor contraction of $C$ and $A^+$,
\begin{align}
\cT_{\j_{x,-1}\j_{x+\hat1,1}\clf_{\cx,-2}\clf_{\cx,2}} 
&= \sum_{\j_{x,2}\j_{x+\hat1,2}} 
A^+_{\j_{x,2}\j_{x+\hat1,2}\clf_{\cx,2}}
C_{\j_{x,-1}\clf_{\cx,-2}\,|\,\j_{x+\hat1,1}\j_{x,2}\j_{x+\hat1,2}} 
\label{Tnew}
\end{align}
with $\Mem \propto D^4$ and $\Comp \propto D^6$.
The total computational cost of the GHOTRG method scales as $D^7$.

\subsection{Staggered phase}
\label{app:staggered}

To avoid multiple definitions of the initial local tensor that would just differ in the staggered phase, we decided not to include the latter in the tensor \eqref{Tm}, but instead to modify the construction of $U$ and the truncation $\cT$ of the coarse tensor, described in the sections above, to take into account the staggered phase.
If we choose the very first contraction to be in the \dir1-direction, these modifications only have to be applied in this contraction. In subsequent contractions, all local tensors on the coarse lattice are identical, and the constructions of $U$ and $\cT$ are performed as described above, without any modifications.
Therefore, to take into account the staggered phase, we only need to modify the equations for the first contraction, by replacing 
$T^{(x)}$ by $\eta_{x,2}^{f_{x,2}}T^{(x)}$ and $T^{(x+\hat1)}$ by $\eta_{x+\hat1,2}^{f_{x+\hat1,2}}T^{(x+\hat1)}$, see also \eqref{Tm}. 
From the definition of the staggered phase, these two staggered phases will have opposite signs. We choose $\eta_{x,2}=1$ and $\eta_{x+\hat1,2}=-1$, such that only operations involving $T^{(x+\hat1)}$ will be affected.

In the calculation of the backward $Q^-$, the non-trivial staggered phase occurs in \eqref{Bb}, but, as it always occurs twice and multiplies to 1, the matrix $Q^-$ remains unaltered. For the forward $Q^+$, the only effect of the staggered phase is to multiply the entries of $Q^+_{(\j_{x,2},\j_{x+\hat1,2}),(\jp_{x,2},\jp_{x+\hat1,2})}$ with $(-1)^{f_{x+\hat1,2}+f'_{x+\hat1,2}}$ in \eqref{Qf}.

In the construction of $\cT$ one just needs to modify \eqref{Tnew} as
\begin{align}
\bar T_{\j_{x,-1}\j_{x+\hat1,1}\clf_{\cx,-2}\clf_{\cx,2}} 
&= \sum_{\j_{x,2}\j_{x+\hat1,2}} (-1)^{f_{x+\hat1,2}}
A^+_{\j_{x,2}\j_{x+\hat1,2}\clf_{\cx,2}}
C_{\j_{x,-1}\clf_{\cx,-2}\,|\,\j_{x+\hat1,1}\j_{x,2}\j_{x+\hat1,2}} .
\end{align}

\section{Analytical results}
\label{app:analytic}

We developed a code to generate all configurations with baryon and meson loops automatically, from which an exact analytic formula for lattices up to $8\times4$ can be computed. The code can be restricted to baryon-loop only configurations or applied to the full meson-baryon system, with chemical potential $\mu$, mass $m$, anisotropy parameter $\gamma$, and arbitrary combinations of periodic and antiperiodic boundary conditions. The results presented below use periodic boundary conditions in space and antiperiodic boundary conditions in time.  Although we derived the analytical results for arbitrary $\gamma$, we only give the formulas for $\gamma=1$ for conciseness.

\subsection{Baryon-only system}

For validation purposes we first computed the partition function in the case where all mesonic contributions are omitted, i.e.,  we only consider configurations where each lattice site is part of a baryon loop. 

The partition functions $Z_B^{L_1\times L_2}$ are given by:
\begin{align}
Z_B^{2\times2} &= 2 \cosh (12 \mu )+6 ,
\label{Z2x2bar}
\\[2mm]
Z^{2\times4}_B &= 98+128\cosh(6\mu)+64\cosh(12\mu)+16\cosh(18\mu)+2\cosh(24\mu) ,
\label{Z2x4bar}
\\[2mm]
Z^{4\times2}_B &= 18 + 2\cosh(24\mu) ,
\label{Z4x2bar}
\\[2mm]
Z^{4\times4}_B &= 2 (839+ 552 \cosh (12 \mu )+360 \cosh (24 \mu )+32 \cosh (36 \mu )+\cosh (48 \mu )) ,
\label{Z4x4bar}
\\[2mm]
Z^{4\times8}_B &= 2 \Big(4887399+7030608 \cosh (12 \mu )+3442496 \cosh (24 \mu )+914544 \cosh (36 \mu )
   \notag\\
   &\hspace{4ex} +154560 \cosh   (48 \mu ) +16416 \cosh (60 \mu )+1152 \cosh (72 \mu )+48 \cosh (84 \mu )+\cosh (96 \mu)\Big) ,
\label{Z4x8bar}
\\[2mm]
Z^{8\times4}_B &= 2 \Big(480087 + 59592 \cosh (24 \mu )+104008 \cosh (48 \mu )+512 \cosh (72 \mu )+\cosh (96 \mu
   )\Big)
.
\label{Z8x4bar}
\end{align}

\subsection{Meson-baryon system}

\allowdisplaybreaks

For the full meson-baryon system the partition functions $Z^{L_1\times L_2}$ are given by:
\begin{align}
Z^{2\times2} &= \frac{998}{9} + \frac{9760}{3} m^2 + 21248 \, m^4 + 50944 \, m^6 + 53248 \, m^8 + 
 24576 \, m^{10} + 4096 \, m^{12} \notag\\
 &+ (16 + 160 \, m^2 + 384 \, m^4 + 256 \, m^6) \cosh 6\mu + 2 \cosh 12\mu ,
\label{Z2x2}
\\[3mm]
Z^{2\times4} &=
\frac{145726}{27}+\frac{12844480}{27} m^2+\frac{105352192}{9} m^4+\frac{3369095168
   }{27}m^6+\frac{2102610944}{3} m^8+2306392064 m^{10}
   \notag\\
   &+\frac{42560880640}{9}m^{12}+6248988672 m^{14}+5386534912 m^{16}+2998927360 m^{18}+1035993088
   m^{20}
   \notag\\
   &+201326592 m^{22}+16777216 m^{24}
   +\cosh (6
   \mu )\bigg(\frac{78640}{27}+\frac{390592}{3}m^2+\frac{5499904}{3} m^4+\frac{103747072}{9} m^6
   \notag\\
   &+38006784 m^8+70975488 m^{10}+77135872 m^{12}+47972352 m^{14}+15728640 m^{16}+2097152 m^{18}\bigg) 
   \notag\\
   &+ \cosh (12 \mu )\left(\frac{4616}{9}+10560 m^2+73856 m^4+217088 m^6+301056 m^8+196608
   m^{10}+49152 m^{12}\right)
   \notag\\
   &+\cosh (18 \mu )\left(48+320 m^2+768 m^4+512 m^6\right)
   +2 \cosh (24 \mu ),
\label{Z2x4}
\\[3mm]
Z^{4\times2} &= \frac{26690}{9} 
   +\frac{9622912}{27}  m^2
   +\frac{89571328}{9}  m^4
   +\frac{3064475648}{27}  m^6
   +\frac{1989659648}{3} m^8
   +2235662336 m^{10} 
   \notag\\&
   +\frac{41867247616}{9} m^{12}
   +6201016320 m^{14}
   +5370806272 m^{16}
   +2996830208 m^{18}
   +1035993088 m^{20}
   \notag\\&
   +201326592 m^{22}
   +16777216 m^{24}
   +\cosh (12 \mu )\big(
   16
   +640 m^2
   +6784 m^4
   +28672 m^6
   +55296 m^8
   \notag\\&
   +49152 m^{10}
   +16384 m^{12}
   \big) 
   +2 \cosh (24 \mu ) ,
\label{Z4x2}
\\[3mm]
Z^{4\times4} &=
\frac{13925769446}{6561}+\frac{680074265344}{729}m^2+\frac{73852306899968}{729}m^4+\frac{3475648217397248}{729}m^6
\notag\\&
+\frac{29728645213136896}{243}m^8
+\frac{471240996681187328}{243}m^{10}
+\frac{14965969111886135296}{729}m^{12}
\notag\\&
+\frac{4131351942668484608}{27}m^{14}
+\frac{2501251147949932544}{3}m^{16}+\frac{276569408916244398080}{81}m^{18}
\notag\\&
+\frac{289682146434549284864}{27}m^{20}
+\frac{236434435268348477440}{9}m^{22}+\frac{4106419252484907204608}{81}m^{24}
\notag\\&
+\frac{699458907104186728448}{9}m^{26}
+95101389643184078848 m^{28}
+\frac{837594161470930681856}{9}m^{30}
\notag\\&
+72746741263060959232 m^{32}
+ 45210379367626047488 m^{34}
+22142023984025174016 m^{36}
\notag\\&
+8422927571833847808 m^{38}
+2432946553384599552 m^{40}
+514817732403789824 m^{42}
\notag\\&
+75153818781745152  m^{44}+6755399441055744 m^{46}
   +281474976710656 m^{48}
\notag\\&
   +\cosh (12 \mu
   )\Biggl(\frac{57524320}{729}+\frac{1295121920}{81} m^2
   +\frac{77688855808}{81} m^4+\frac{2123443054592}{81} m^6
\notag\\&
   +\frac{32295371456512}{81} m^8
   +\frac{101785491374080}{27} m^{10}
   +\frac{1923114590765056}{81} m^{12}+\frac{938298391396352}{9} m^{14}
\notag\\&
   +\frac{987744196100096}{3} m^{16}+\frac{6871321849888768}{9} m^{18}
   +1316708023271424 m^{20}+\frac{5097380869308416}{3} m^{22}
\notag\\&
   +1639786838228992 m^{24}
   +1174310630719488 m^{26}+613377164443648 m^{28}+226465035583488 m^{30}
\notag\\&
   +55868934586368 m^{32}+8246337208320 m^{34}+549755813888 m^{36}\Biggr) 
   +\cosh (24 \mu )\Biggl(\frac{421736}{81}+\frac{2784512}{9} m^2 
\notag\\&
   +\frac{24326656}{3} m^4+\frac{914487296}{9} m^6+703141888 m^8+2929164288 m^{10}
   +7757299712 m^{12}
\notag\\&
   +13481541632 m^{14}
   +15568732160 m^{16}+11853103104 m^{18}+5737807872 m^{20}
   +1610612736 m^{22}
\notag\\&
   +201326592 m^{24}\Biggr) 
   +\cosh (36 \mu )\Bigl(160+2048 m^2+16640 m^4+59392
   m^6+110592 m^8+98304 m^{10}
\notag\\&
   +32768 m^{12}\Bigr) 
   +2 \cosh (48 \mu )  .
\label{Z4x4}
\end{align}

\bibliographystyle{elsarticle-num.bst}
\bibliography{biblio} 

\end{document}